\documentclass[12pt]{article}
\pdfoutput=1
\usepackage{color,amsmath,amssymb,exscale,psfrag,epsfig}
\usepackage{cite,color,url}
\usepackage{pdfpages}	
\usepackage{graphicx}
\usepackage{slashed}
\usepackage[utf8]{inputenc}
\usepackage[T1]{fontenc}
\usepackage{xcolor}
\usepackage{accents}
\usepackage{centernot}
\usepackage{fixltx2e}
\usepackage{enumerate}
\usepackage{subfig}
\usepackage{float}
\usepackage{floatpag}
\usepackage{placeins}
\floatpagestyle{empty}

\newcommand{\met}{\ensuremath{E_{\textrm{T}}^{\textrm{miss}}}}

\newcommand{\hthad}{\ensuremath{H_{\textrm{T}}}}

\newcommand{\fourtop}{\ensuremath{t\bar{t}t\bar{t}}}

\usepackage{lineno}

\def\bea{\begin{eqnarray}}
\def\eea{\end{eqnarray}}

\definecolor{nicered}{rgb}{0.7,0.1,0.1}
\definecolor{nicegreen}{rgb}{0.1,0.5,0.1}
\newcommand{\Att}{A^{L/U}_\textrm{tt-hel}}

\catcode`\@=11
\def\lsim{\mathrel{\mathpalette\@versim<}}
\def\gsim{\mathrel{\mathpalette\@versim>}}
\def\@versim#1#2{\vcenter{\offinterlineskip
\ialign{$\m@th#1\hfil##\hfil$\crcr#2\crcr\sim\crcr } }}
\catcode`\@=12
\catcode`@=12
\begin{document}
\thispagestyle{empty}
\begin{flushright}
ICAS 043/19
\end{flushright}
\vspace{0.1in}
\begin{center}
{\Large \bf Four-top as probe of light top-philic New Physics} \\
\vspace{0.2in}
	{\bf Ezequiel \'Alvarez$^{(a)\dagger}$,
	Aurelio Juste$^{(b,c)\star}$,
	Rosa Mar\'ia Sand\'a Seoane$^{(a),\diamond}$
}
\vspace{0.2in} \\
	{\sl $^{(a)}$ International Center for Advanced Studies (ICAS)\\
 UNSAM, Campus Miguelete, 25 de Mayo y Francia, (1650) Buenos Aires, Argentina }
\\[1ex]
{\sl $^{(b)}$
Institut de F\'isica d'Altes Energies (IFAE), Edifici Cn, Facultat de Ci\`encies,\\
Universitat Aut\`onoma de Barcelona, E-08193 Bellaterra, Barcelona, Spain}
\\[1ex]
{\sl $^{(c)}$
Instituci\'o Catalana de Recerca i Estudis Avan\c{c}ats (ICREA), E-08010 Barcelona, Spain}
\end{center}
\vspace{0.1in}

\begin{abstract}
We study the four-top ($t\bar t t \bar t$) final state at the LHC as a probe for New Physics (NP)  effects due to new particles that couple predominantly to the top quark and whose masses are below the top-quark-pair production threshold.  We consider simple NP models containing a new particle with either spin 0, spin 1, or spin 2, and find benchmark points compatible with current experimental results.  We find that interference effects between NP and QED amplitudes can be large, pointing out the necessity of NLO contributions to be explicitly computed and taken into account when NP is present.  We examine kinematic differences between these models and the Standard Model (SM) at the parton level and the reconstructed level. In the latter case, we focus on events selected requiring two same-sign leptons and multiple jets. We investigate how the different Lorentz structure of the light NP affects the kinematic hardness, the polarization, the spin correlations, and the angular distributions of the parton-level and/or final-state particles.  We find that spin-2 light NP would be identified by harder kinematics than the SM.  We also show that the angular separation between the same-sign leptons is a sensitive observable for spin-0 NP.  The spin-0 and spin-2 NP cases would also yield a signal in $t\bar t \gamma\gamma$ with the invariant mass of the photons indicating the mass of the new particle.  The spin-1 NP would be identified through an excess in four-top signal and slight or not modification in other observables, as for instance the lack of signal in $t\bar t \gamma\gamma$ due to the Landau-Yang theorem.  We comment on the opportunities that would open from the kinematic reconstruction of some of the top quarks in the $t\bar t t \bar t$ state. Our results provide new handles to probe for light top-philic NP as part of the ongoing experimental program of searches for four-top production at the LHC Run 2 and beyond.
\end{abstract}

\vspace*{2mm}
\noindent {\footnotesize E-mail:
{\tt 
$\dagger$ sequi@unsam.edu.ar,
$\star$ juste@ifae.es,
$\diamond$ rsanda@unsam.edu.ar.
}
}

\newpage
\section{Introduction}
\label{section:1}

The LHC is already a very successful machine. It has discovered a Standard Model (SM)-like Higgs boson \cite{Aad:2012tfa,Chatrchyan:2012xdj}, which was one of the main drivers of its design and construction, and it has pushed our frontiers of knowledge to extraordinary limits by excluding the existence of new particles over a broad range of masses and couplings in a wide variety of New Physics (NP) models.   The current state-of-the-art in High Energy Physics (HEP) research can be depicted as a vast and challenging ocean, of which we are practically clueless, between the current TeV energy frontier and the Planck energy scale.  Over the next two decades, while the LHC completes its Run 3, the HEP community will be devoted to the scrutiny of all available LHC results, as well as to the proposal of new promising experimental directions.  Among these upcoming LHC results, there are few processes that are beginning to be tested experimentally using the full Run 2 dataset, and whose measurement is directly sensitive to NP contributions. Of particular interest are Higgs-boson pair ($hh$) production~\cite{DiMicco:2019ngk,Aad:2019yxi,CMS:2019vgr}, the associated production of a Higgs boson with a top-antitop-quark pair ($t\bar t h$)~\cite{CMS:2018rbc,Madaffari:2018bbq,Aaboud:2018urx}, and four-top ($t\bar t t \bar t$) production \cite{Aad:2015kqa,Aad:2015gdg,Sirunyan:2019wxt}. The first two processes, $hh$ and $t\bar t h$, will deliver crucial direct information on the Higgs potential and the top-quark Yukawa coupling, respectively.  The latter process, $t\bar t t \bar t$, can also be used to probe the top-quark Yukawa coupling, the Higgs-boson width~\cite{Cao:2016wib,Cao:2019ygh}, and anomalous off-shell Higgs behavior~\cite{Englert:2019zmt}.  In addition, it has unique sensitivity to light top-philic NP, and thus represents an exciting opportunity for discovery at the LHC. This is our primary motivation to study it in this work.

Theory and phenomenology works considering $t\bar t t \bar t$ as a sensitive final state for top-philic NP can be found in Refs.~\cite{Alvarez:2016nrz,Battaglia:2010xq,Degrande:2010kt,Alvarez:2017wwr,Lillie:2007hd,Greiner:2014qna,Azzi:2019yne,Calvet:2012rk,Darme:2018dvz,Zhang:2017mls,Alvarez:2019knh}.  Most of these articles consider heavy NP, above the $t\bar{t}$ production threshold.   Searches for SM four-top production have been performed by the ATLAS and CMS collaborations at center-of-mass energies of 
8 TeV~\cite{Aad:2014pda,Khachatryan:2014sca, Aad:2015kqa,Aad:2015gdg} and 13 TeV~\cite{Sirunyan:2017tep,Sirunyan:2017uyt, Sirunyan:2017roi,Aaboud:2018jsj,Aaboud:2018xpj,Sirunyan:2019nxl,Sirunyan:2019wxt}. These searches have focused on either the single-lepton (1L) and opposite-sign dilepton (2LOS) channels,
or the same-sign dileptons (2LSS) and multilepton (ML) channels. In all cases the final states signature is spectacular, featuring in addition 
to the leptons, a high multiplicity of jets, four of which originate from the hadronization of $b$-quarks ($b$-jets). Although the four-top signal yield
is highest in the 1L and 2LOS channels, these searches are extremely challenging due to the overwhelming background from 
$t\bar{t}$ production in association with heavy-flavor jets, which suffers from large uncertainties in its theoretical modeling. In contrast,
searches in the 2LSS and ML channels have lower signal yield, but also much more manageable backgrounds. The main backgrounds primarily originate from $t\bar{t}W$, $t\bar{t}(Z/\gamma^*)$, and $t\bar{t}H$ production, as well as from $t\bar{t}$  production with additional leptons from 
heavy-flavour hadron decays, misidentified jets, or photon conversions, and other processes where the electron charge is incorrectly assigned.
The most sensitive search for SM four-top production to date has been performed by the CMS Collaboration considering the 
2LSS and ML channels, and using the full Run 2 dataset, corresponding to 137~fb$^{-1}$ of integrated luminosity at $\sqrt{s}=13$~TeV~\cite{Sirunyan:2019wxt}.
The observed (expected) significance for the SM four-top signal is 2.6 (2.7) standard deviations (s.d.), and the measured value of the
SM four-top cross-section is $\sigma^\textrm{meas}_{\fourtop} = 12.6^{+5.8}_{-5.2}$~fb, in agreement with the SM prediction of 
$\sigma^\textrm{SM}_{\fourtop} = 12.0^{+2.2}_{-2.5}$~fb~\cite{Frederix:2017wme}, which includes NLO QCD and electroweak effects. 
The resulting ratio between the measured and predicted cross-sections is
$\mu_{\fourtop} = \sigma^\textrm{meas}_{\fourtop} / \sigma^\textrm{SM}_{\fourtop}  = 1.05^{+0.52}_{-0.48}$. Therefore, an enhancement in the four-top production cross-section due to NP contributions of up to a factor of 1.5 (2.0) is still compatible with 
the measurements at about 1 s.d. (2 s.d.) level.

This article is organized as follows.  In Sect.~\ref{section:2} we describe a set of simple NP models containing new particles with either spin 0, spin 1, or spin 2, which couple predominantly to the top quark and whose masses are below the $t\bar{t}$ production threshold.  We study existing constraints in these NP models arising from searches for di-photon ($\gamma\gamma$) resonances and for four-top production, and we define suitable benchmark points in the model parameter space compatible with experimental results.  In Sect.~\ref{section:3} we study the phenomenology of four-top production for some of these benchmark points.  In Sect.~\ref{section:4} we present a discussion on the obtained results, and Sect.~\ref{section:5} contains the conclusions.  We include three appendices to show more results on the full set of benchmark points, to describe the one-loop features of the NP models, and to summarize the numerical simulation details.

\section{NP Models}
\label{section:2}

The aim of this article is to study how simple NP models would affect the four-top phenomenology at the LHC, and how could they be recognized and distinguished.  We focus on models whose effects are expected to be more important in four-top production rather than in other processes.  To this end, we consider new particles whose couplings are predominantly to the top quark and whose mass $M$ is below the $t\bar t$ threshold ($M < 2 m_t$) to avoid resonance effects in $t\bar{t}$ production. In addition, we restrict ourselves to new particles which are color neutral to avoid interactions with gluons that would yield a large QCD-mediated production cross-section.  For this purpose we study the following simple models that are described below: {\it i)} Scalar, {\it ii)} Pseudo-scalar, {\it iii)} vector $Z'$,  and {\it iv)} Graviton.

It is not the objective of this work to develop the UV completion of the proposed simple NP models. However, it can be argued that having a new resonance with couplings to SM particles dominated by the top quark is feasible, as for instance in two-sector models~\cite{Contino:2006nn} or Composite Higgs Models (CHM) \cite{Marzocca:2012zn}.  In these models the SM is accompanied by a heavier strongly interacting sector; the details and phenomenology of this kind of NP can be found elsewhere \cite{Contino:2006nn}.  In CHM models, to avoid experimental constraints and to explain the fermion mass hierarchy, it is customary to implement partial compositeness, where the degree of compositeness of each physical fermion depends on its mass \cite{Caracciolo:2012je}. Given Electroweak Precision Tests on $SU(2)_L$ it is convenient to set in the model the right-chiral top quark ($t_R$) with large compositeness \cite{Contino:2006nn}.  Then, depending on the particular realization of each model, one can obtain for different cases some new light resonances that couple predominantly to $t_R$.  For the case of a spin-0 field, however, the Lorentz structure requires a left-chiral top quark ($t_L$) as well.  Vertices of this kind are found in Ref.\cite{Alvarez:2016nrz}, and Ref.~\cite{Dev:2014yca} discusses a UV complete model with additional scalars that presents this kind of phenomenology.

For the sake of simplicity, and to address qualitative aspects of four-top production phenomenology, throughout this work we make the assumption that the new light resonance couples only to the top quark, as described in the following paragraphs.

\subsection{NP Interaction Lagrangian}

\noindent{\bf Scalar NP: $\phi$}

For the scalar case we study the following simplified Lagrangian
\begin{equation}
{\cal L}_{\phi}^\textrm{tree} = g_{\phi t} \, \bar t_L \phi t_R  + h.c. \ .
\end{equation}

A one-loop effective coupling to gluons ($\phi gg$) and to photons ($\phi \gamma\gamma$) is added to the Lagrangian through a top-quark loop. The NP effective interaction Lagrangian therefore reads
\begin{equation}
{\cal L}_{\phi} = {\cal L}_{\phi}^\textrm{tree} + {\cal L}_{\phi g g}^\textrm{eff}+{\cal L}_{\phi \gamma \gamma}^\textrm{eff} \ .
\end{equation}
Details on the one-loop effective Lagrangian are described in App.~\ref{loop}.

\noindent{\bf Pseudo-scalar NP: $A$}

The pseudo-scalar case has the following Lagrangian
\begin{equation}
{\cal L}_{A}^\textrm{tree} = g_{A t} \, \bar t_L A i \gamma^5 t_R + h.c. \ .
\end{equation}

Including the one-loop effective Lagrangian describing $Agg$ and $A \gamma\gamma$ interactions that can be found in App.~\ref{loop}, the full pseudo-scalar Lagrangian reads 
\begin{equation}
{\cal L}_{A} = {\cal L}_{A}^\textrm{tree} + {\cal L}_{A g g}^\textrm{eff}+{\cal L}_{A \gamma \gamma}^\textrm{eff} \ .
\end{equation}

\noindent{\bf $Z'$ Vector NP: $Z'$}

For the purpose of our work, the only interaction considered for $Z'$ reads
\begin{equation}
{\cal L}_{Z'} =  g_{Z't} \,  Z'_\mu \bar t_R \gamma^\mu t_R \ .
\end{equation}
Since $Z'$ is a spin-1 particle, it cannot couple at any order to a $\gamma\gamma$ final state due to the Landau--Yang theorem~\cite{Landau:1948kw,Yang:1950rg}.  However, recent works claim that a spin-1 particle can couple to a $gg$ state since gluons are colored \cite{Beenakker:2015mra,Pleitez:2018lct,Cacciari:2015ela,Pleitez:2015cpa}.  Since we expect that limits coming from resonance searches in di-jet ($jj$) production and $t\bar{t}$+jets production ($t jj$ or $t\bar t jj$) do not have sufficient sensitive to probe the relevant parameter space for the models considered, we ignore this possibility.

As pointed out in Ref.~\cite{Alvarez:2016nrz}, having the $Z'$ coupled to an unconserved current such as $\bar t_R \gamma^\mu t_R$ yields a factor $(m_t/M)^2$, due to the longitudinal polarization of the vector propagator. For small $M$ this translates into an enhancement in the cross-section, as discussed below.  We have explicitly verified that if we use the conserved fermionic current instead, then this enhancement for small $M$ disappears and the behavior is more similar to the spin-0 case, where the coupling is to a conserved current.

\noindent{\bf Graviton NP: $G$}

We consider an effective Lagrangian for a spin-2 graviton with field $\hat G_{\mu\nu}$.  The tree-level interaction Lagrangian reads \cite{Alvarez:2016ljl}
\begin{equation}
{\cal L}_{G}^\textrm{tree} = - \frac{i}{2 \Lambda}  \hat G ^{\mu\nu} \left[ g_{Gt} \left( \bar t_R \gamma_\mu \overset{\text{\tiny $\leftrightarrow$}}{D}_\nu t_R - \eta_{\mu\nu} \bar t_R \gamma^\rho \overset{\text{\tiny $\leftrightarrow$}}{D}_\rho t_R \right) \right]\ ,
\end{equation}
\noindent where $\bar f \gamma_\mu \overset{\text{\tiny $\leftrightarrow$}}{D}_\nu f = \bar f \gamma_\mu D_\nu f - D_\nu \bar f \gamma_\mu f$. Contrary to the previous NP models, the spin-2 Lagrangian needs dimensional couplings, hence the dimensional constant $\Lambda$ in the denominator.  The constant $\Lambda$ can be understood as the energy scale up to which the theory as described here is valid.  Throughout the remainder of this article we set $\Lambda=3\ \mbox{TeV}$.  

It is interesting to notice that in this model, in addition to the coupling between the resonance and the top-quark pair, SM gauge invariance introduces 4-point interactions that include the resonance, the top-quark pair, and a SM gauge boson.  This represents a distinctive feature of the model, since there are Feynman diagrams in $pp \to t\bar t t \bar t$ production that are not present in the other models (see Fig.~\ref{feynman}c). 

Therefore, the full spin-2 Lagrangian reads 
\begin{equation}
{\cal L}_G = {\cal L}_{G}^\textrm{tree} + {\cal L}_{Ggg}^\textrm{eff}+ {\cal L}_{G\gamma\gamma}^\textrm{eff} \ ,
\end{equation}
\noindent where the one-loop effective Lagrangians due to gluons and photons can be found in App.~\ref{loop}.

\subsection{Constraints on the NP models}
\label{constraints}

Since we restrict our study to cases where the mass of the NP resonance is below the $t\bar{t}$ production threshold, $M \lsim 350$ GeV, the constraints on the model would mainly come from NP loop corrections to $t\bar{t}$ near threshold, $\gamma\gamma$ resonance searches, and four-top production at the LHC.  We examine these constraints in the following paragraphs. 

Loop corrections to $t\bar{t}$ production have been studied for Higgs and electroweak gauge bosons \cite{Kuhn:2013zoa}.  Although it is not possible to directly extract bounds on the presented NP models from the available results, adapting these computations for spin-0 and spin-1 NP contributions, as well as including spin-2 corrections and the corresponding interference with SM contributions, could provide relevant constraints on the models.  This objective lies beyond the scope of this work; however, given the precision reached in $t\bar{t}$ production and its recent application in constraining the top-quark Yukawa coupling~\cite{CMS:2019unu}, we estimate that results near the $t\bar{t}$ threshold should be interesting concerning the presented NP models.

\begin{figure}[hp!]
\begin{center}
\subfloat[]{\includegraphics[width=0.40\textwidth]{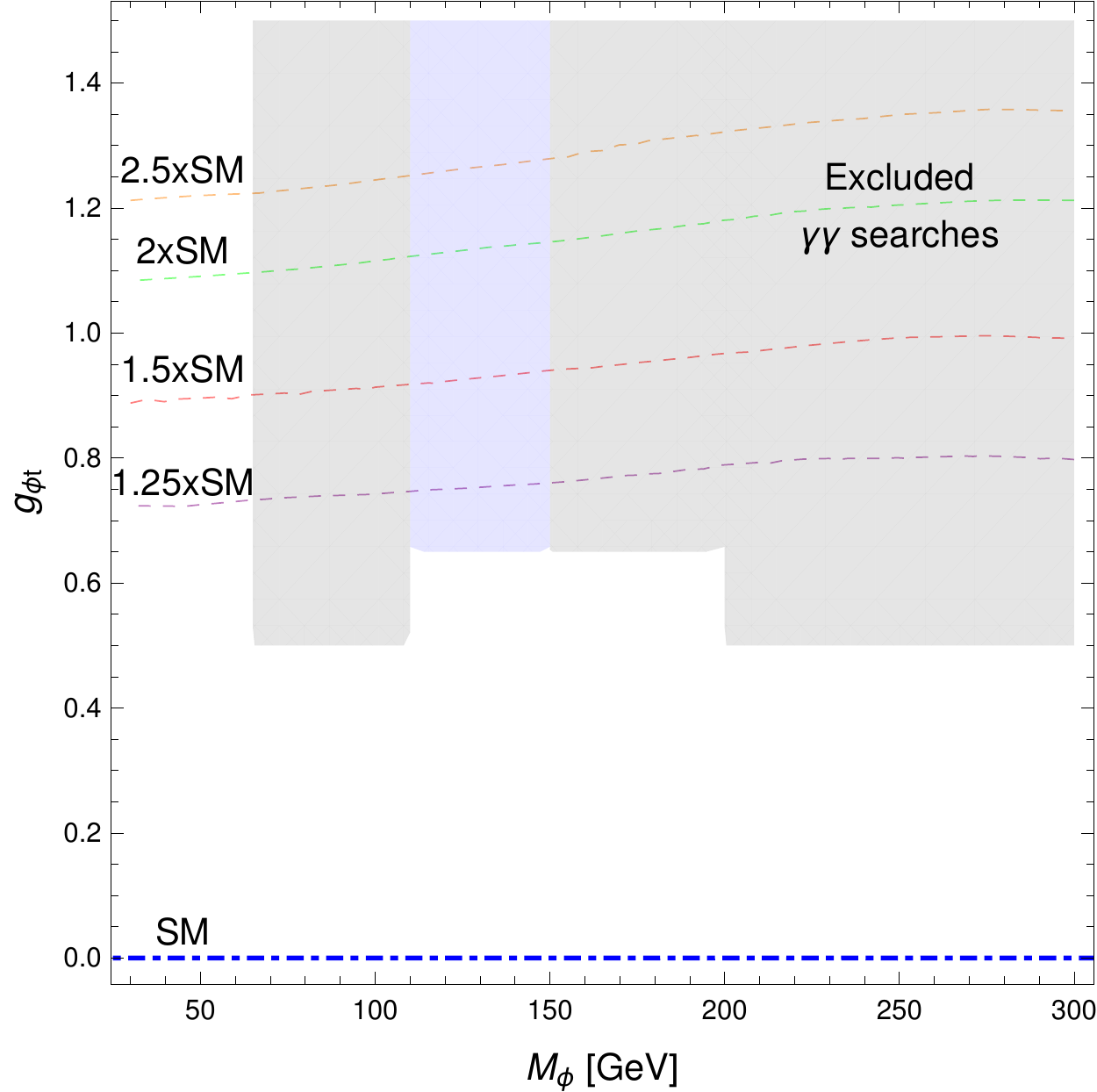}}\hspace{3mm}
\subfloat[]{\includegraphics[width=0.40\textwidth]{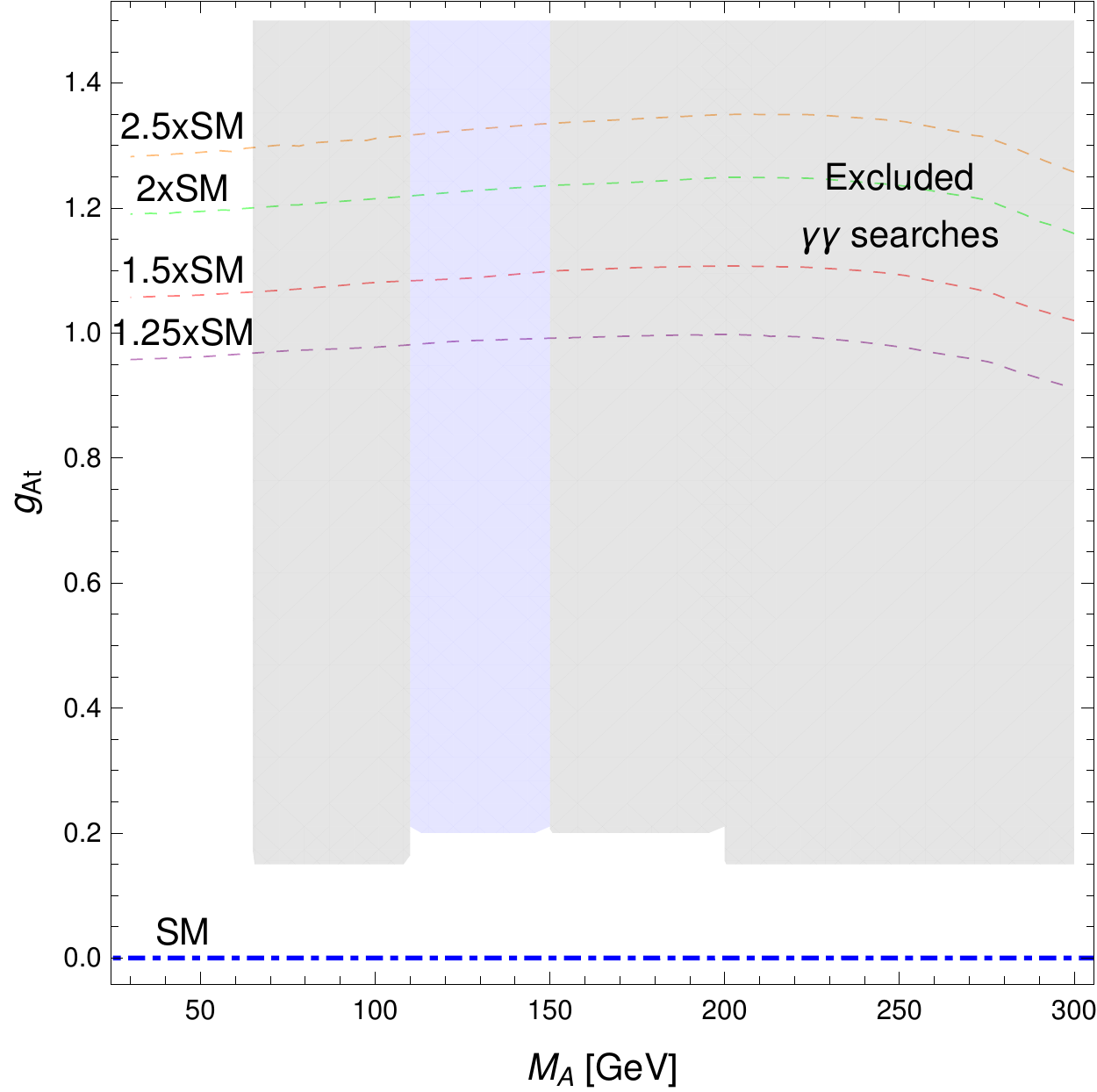}} \\
\subfloat[]{\includegraphics[width=0.40\textwidth]{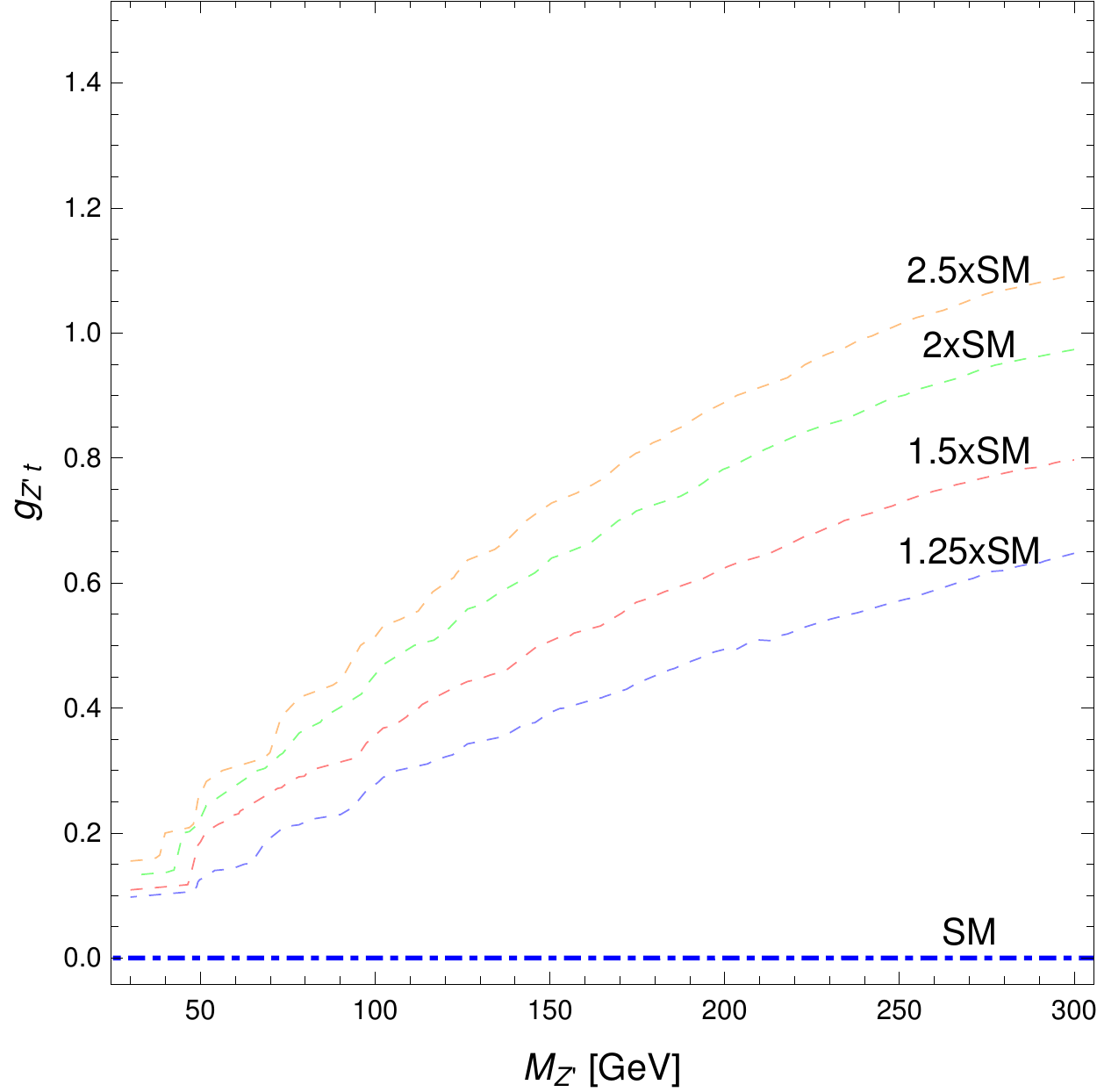}}\hspace{3mm}
\subfloat[]{\includegraphics[width=0.40\textwidth]{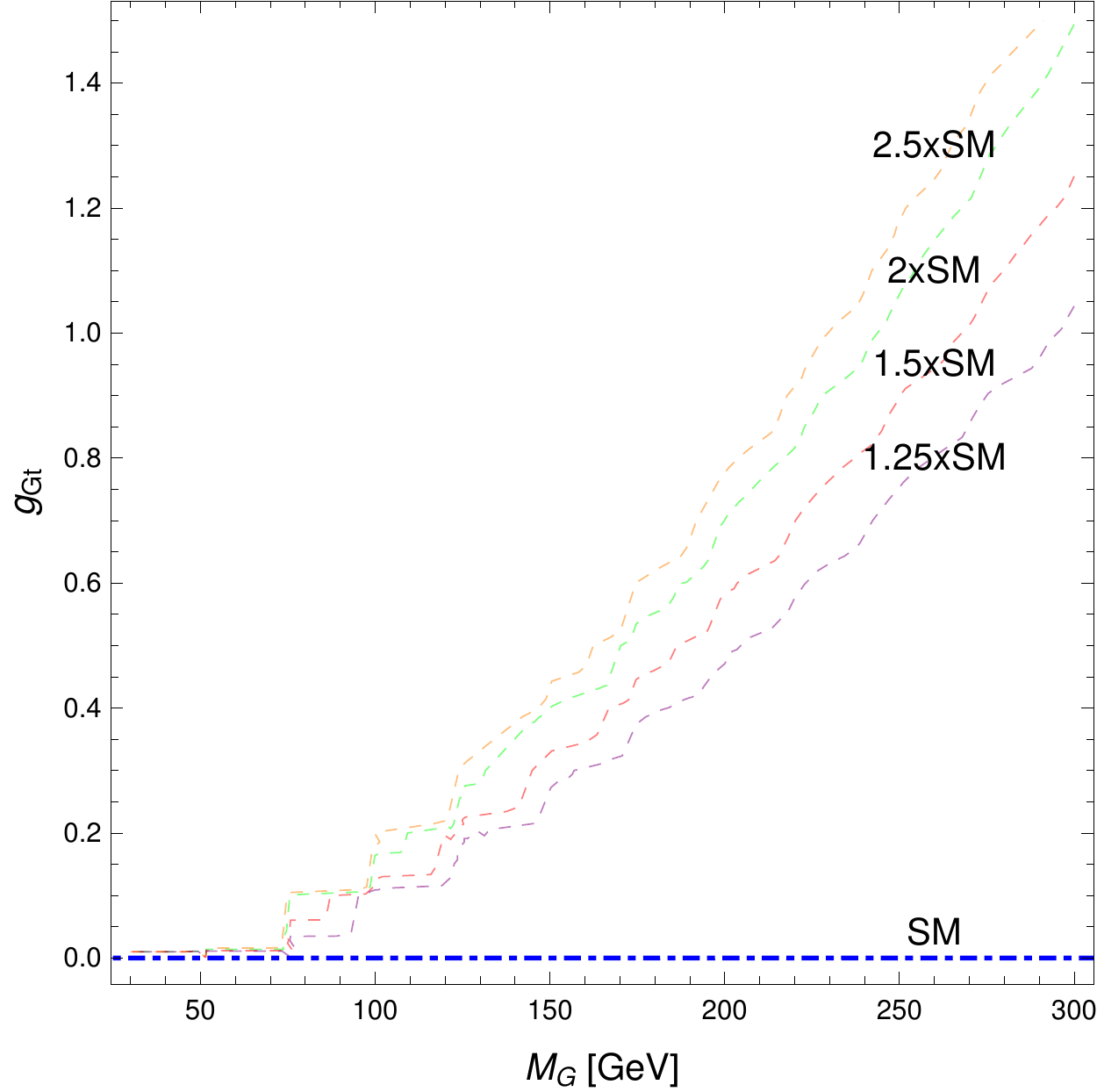}} \\
	\caption{\small Four-top production cross-section contour levels (in units of the SM cross-section) in the plane of coupling strength versus resonance mass for the different NP models considered; (a) Scalar, (b) Pseudo-scalar, (c) $Z'$ and (d) Graviton.  The $Z'$ and Graviton models exhibit a larger four-top production cross-section dependence on the resonance mass since the NP is coupled to an unconserved current, as explained in the text. 
Values of the predicted cross-section above about $2\times$SM are excluded at the 95\% CL by the latest four-top search at the LHC (under the assumption of SM kinematics). The gray shaded area represents the region excluded by $\gamma\gamma$ resonance searches. The light-blue shaded region corresponds to an interpolation since there are not available general $\gamma\gamma$ searches in the 110 GeV - 150 GeV range, where the $H \to \gamma\gamma$ signal is measured and no other excesses are observed.  In the case of the $Z'$ and Graviton models, such $\gamma\gamma$ resonance searches do not yield significant constraints: the $Z'$ resonance cannot decay into $\gamma\gamma$, whereas the  Graviton model considered is less sensitive to $\gamma\gamma$ searches as explained in the text. 
}
\label{bounds1}
\end{center}
\end{figure}

New particles with spin $\neq 1$ can be created in gluon fusion through a top-quark loop and decay into a $\gamma\gamma$ or di-jet final state through a top-quark loop, as discussed previously. A massive spin-1 particle cannot decay to $\gamma\gamma$ due to the Landau-Yang theorem \cite{Landau:1948kw,Yang:1950rg}.  Since parton-level calculations yield a ratio of $S/B$ ($S/\sqrt{B}$) between $\gamma\gamma$ and di-jet final states of $\sim 10^5$ ($\sim 10$), and since there are no updated di-jet resonance searches in the relevant region of invariant masses, we only consider $\gamma\gamma$ resonance searches.   The latter represent an extensive program by both ATLAS \cite{Aad:2014ioa,Aaboud:2017yyg,ATLAS:2018xad} and CMS \cite{Khachatryan:2015qba,Sirunyan:2018aui}.  We have scanned the parameter space of the relevant models and compared the predicted cross-sections with the available experimental bounds. The details of the simulations are described in App.~\ref{appendix}.  We present these bounds in Fig.~\ref{bounds1}.  We find that $\gamma\gamma$ resonance searches provide important bounds for the spin-0 NP model, being more restrictive for the pseudo-scalar case. 
On the other hand, for the spin-2 model, we find that the four-top cross-section is enhanced by the extra Feynman diagrams compared to the other NP scenarios.  Thus for regions in parameter space with same four-top cross-section as in the spin-0 models, the $\gamma\gamma$ process has smaller cross-section in the spin-2 model. As a result the spin-2 model remains rather unconstrained by the available $\gamma\gamma$ resonance searches.

\begin{figure}[h!]
\begin{center}
	\subfloat[]{\includegraphics[width=0.40\textwidth]{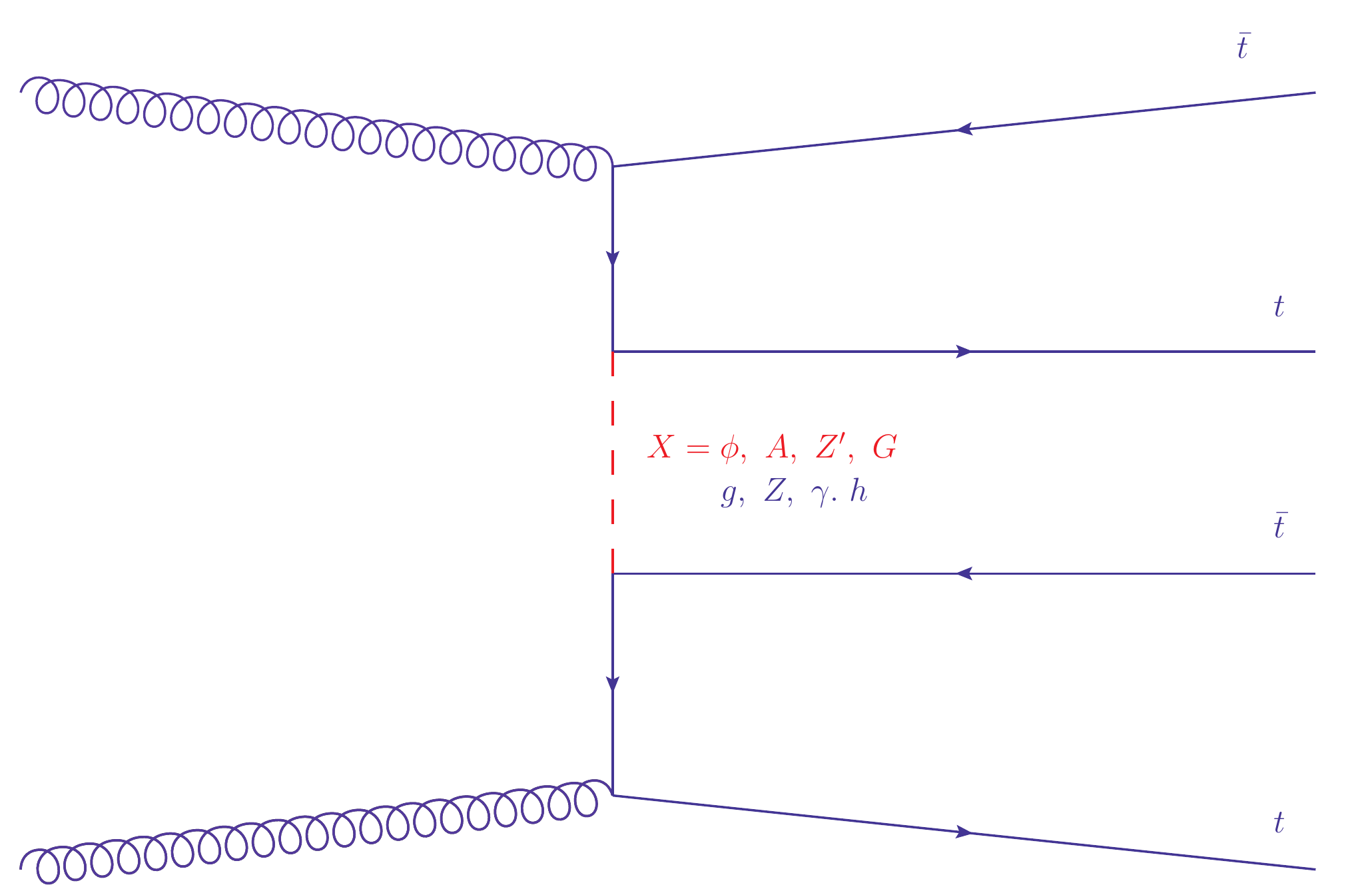}} \hspace{3mm}
	\subfloat[]{\includegraphics[width=0.40\textwidth]{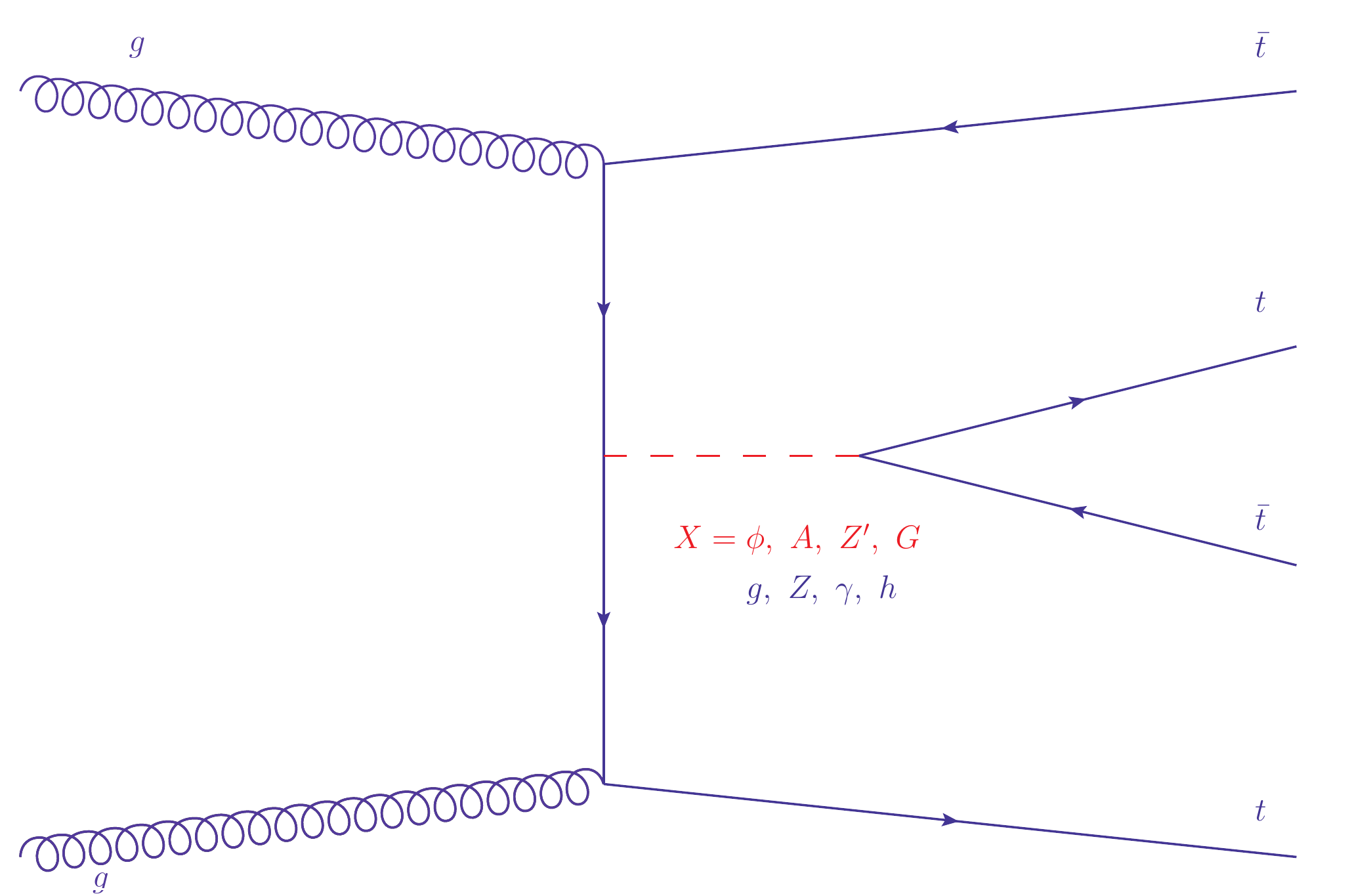}} \hspace{3mm}
	\subfloat[]{\includegraphics[width=0.40\textwidth]{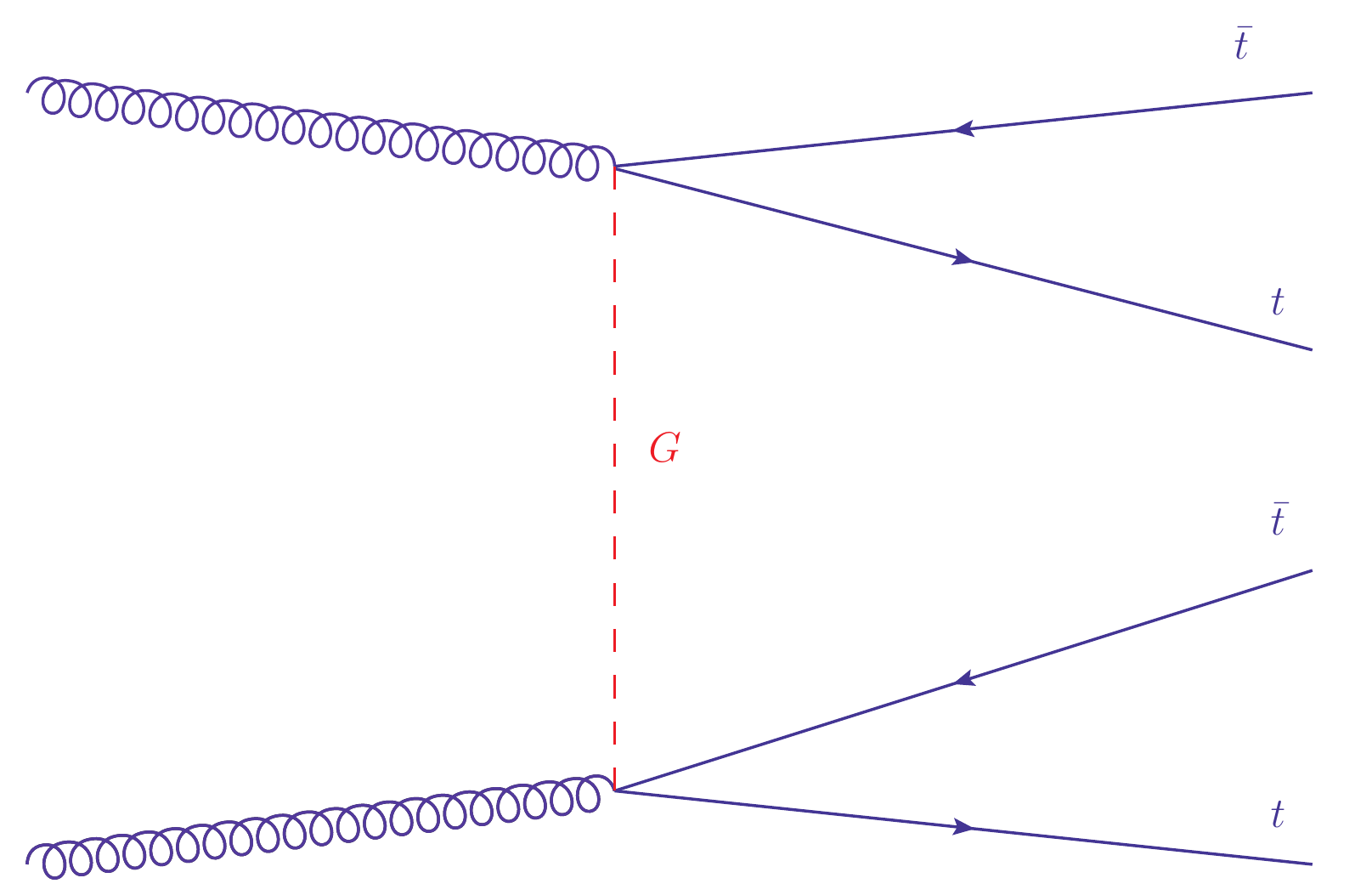}} \hspace{3mm}
	\subfloat[]{\includegraphics[width=0.40\textwidth]{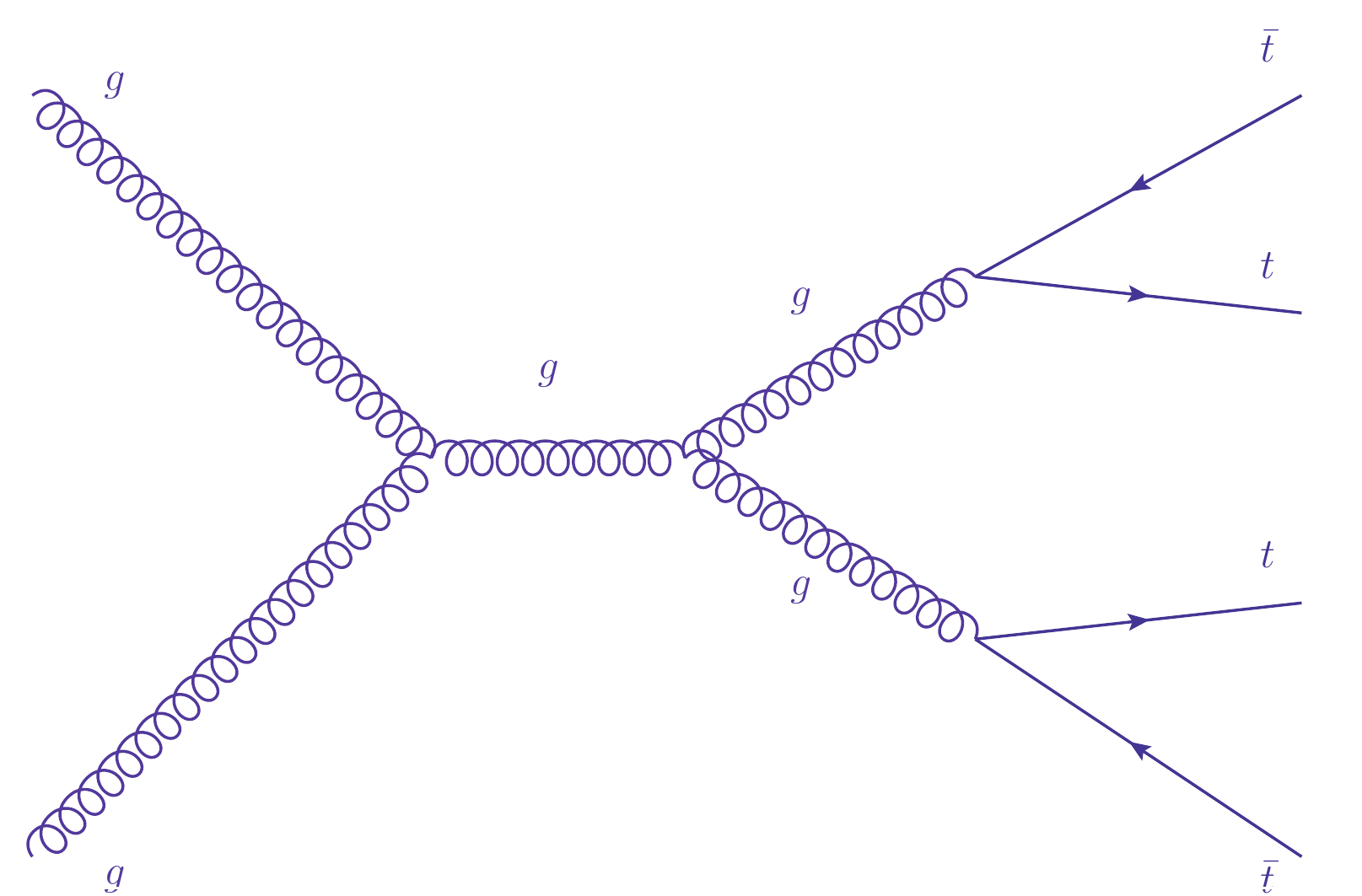}}
	\caption{\small Representative Feynman diagrams for the $pp\to t \bar t t \bar t$ process at the LHC where, at $\sqrt{s}=13$~TeV, approximately 90\% of the production cross-section is through gluon fusion.  In (a) and (b) the dashed line may correspond to any of the NP particles considered in this work ($\phi$, $A$, $Z'$ or $G$), in addition to the SM particles ($g,\ Z,\ \gamma$ and Higgs).  Diagram (c) is only allowed in the spin-2 NP model, where the 4-point interaction is required to ensure gauge invariance. Diagram (d) correspond to a typical QCD SM diagram. Diagrams (a) and (c) are important for not having $s$-channel suppression.  In particular, in diagram (c) the Graviton model provides four-top production through only 2 vertices without $s$-channel suppression and with all four top quarks on an equal footing, which provides distinctive features (see text for discussion).
	}
\label{feynman}
\end{center}
\end{figure}

\begin{figure}[t]
\begin{center}
\subfloat[]{\includegraphics[width=0.30\textwidth]{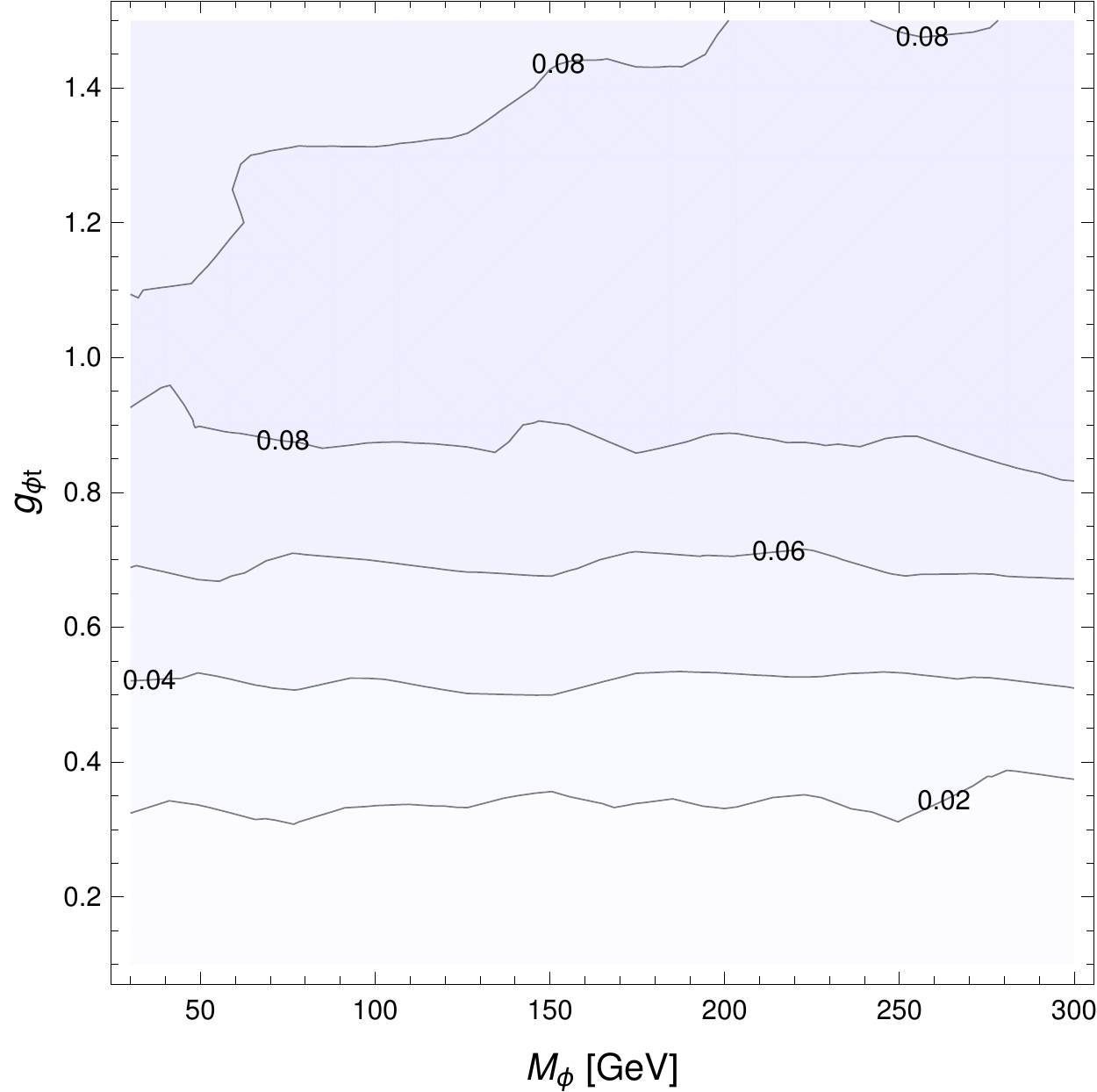}}\hspace{3mm}
\subfloat[]{\includegraphics[width=0.30\textwidth]{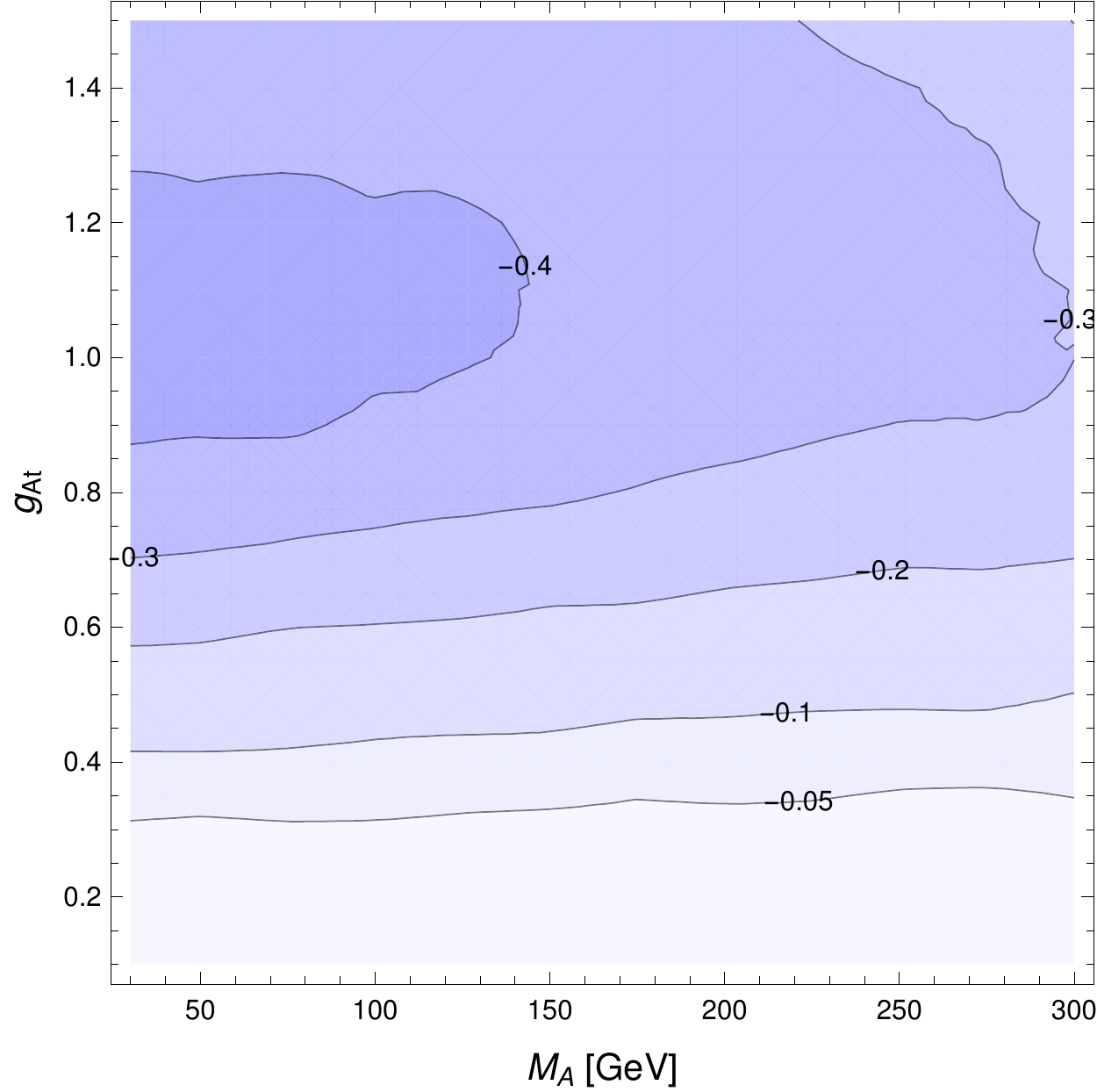}} \hspace{3mm}
\subfloat[]{\includegraphics[width=0.30\textwidth]{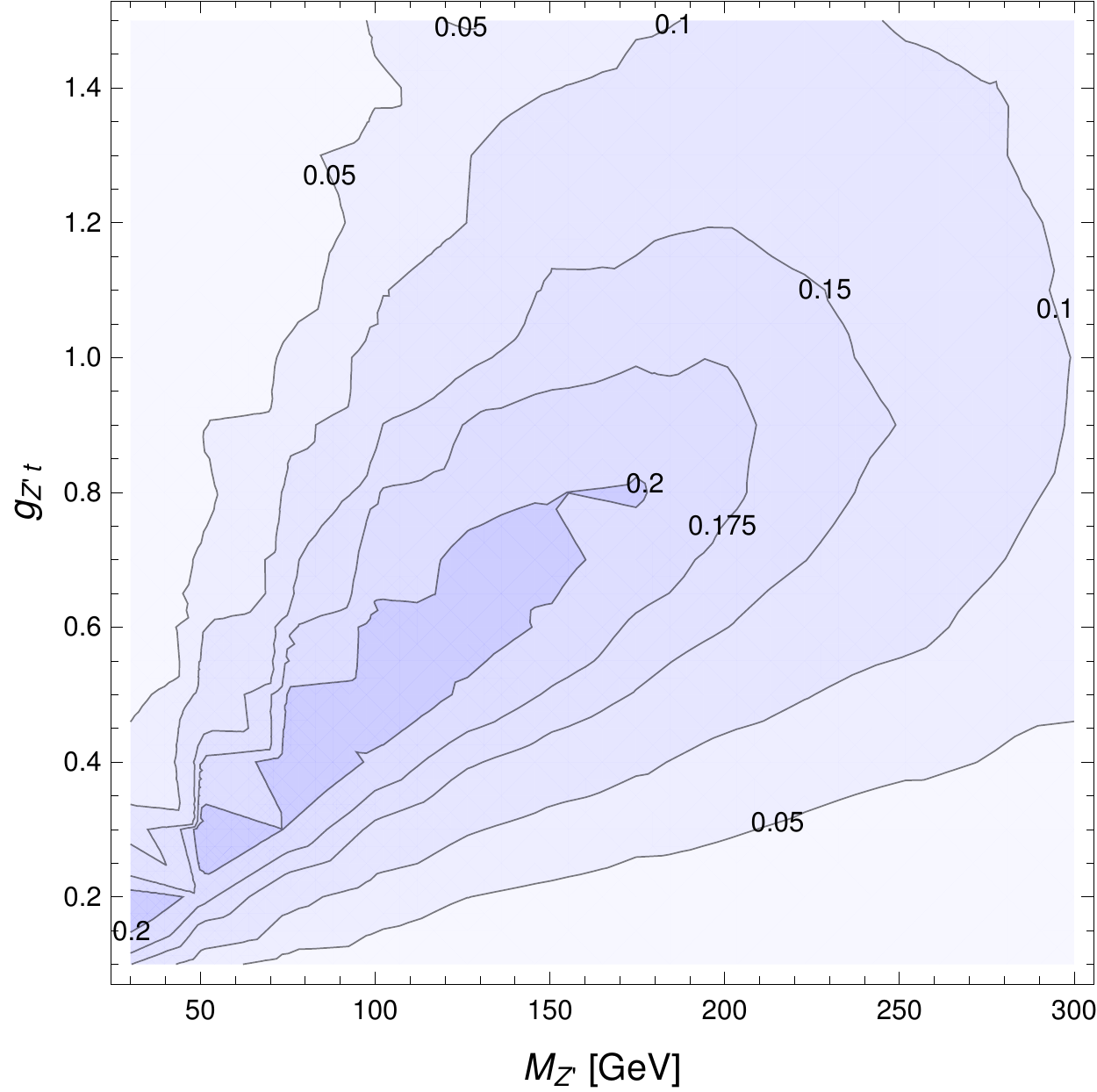}}\\
\caption{\small Fraction of the contribution to the total $pp\to t \bar t t \bar t$ cross-section due to interference between SM (QCD+QED) and NP.  It is important to stress that the interference is enhanced by including QED corrections, as discussed in text.  In the Pseudo-scalar benchmark points the interference may reach up to $\sim -70\%$ of the total SM LO cross-section.  The interference for the Graviton benchmark points --not shown-- is negligible in the relevant region. }
\label{interference}
\end{center}
\end{figure}

Some representative Feynman diagrams for the $pp \to t\bar t t \bar t$ process are presented in Fig.~\ref{feynman}. 
As discussed in Sect.~\ref{section:1}, the most sensitive SM four-top search to date~\cite{Sirunyan:2019wxt} excludes at the 95\% CL values of the four-top cross-section larger than approximately a factor of two larger the SM prediction (under the assumption of SM kinematics), thus still leaving enough room for light top-philic NP contributions.
We show in Fig.~\ref{bounds1} the contour-levels of the predicted SM+NP cross-sections in units of the SM cross-section.  Simulations are in equal conditions for SM and NP, which is equivalent to using the same NLO k-factor for both scenarios.  Details on the simulation process are given in App.~\ref{appendix}.  It is interesting to notice that, although LO electroweak corrections represent a minor correction of $\sim 5 \%$ to the SM cross-section, their fractional contribution is enhanced when NP effects are included.  
 This could be expected, since it has been shown in Ref.~\cite{Frederix:2017wme} that SM LO contributions of $\mathcal{O}(\alpha_s^3 \alpha)$ and $\mathcal{O}(\alpha_s^2 \alpha^2)$ are both sizable but have opposite sign, leading to a large accidental cancellation.   Therefore, any NP contribution that affects the interference terms may break this cancellation at this order, thus resulting in larger contributions to the total cross-section.  As stated in Ref.~\cite{Frederix:2017wme}, a similar behavior is also expected at NLO, where there is also a cancellation between SM contributions of different order. In general we find that this enhancement at LO is due to the interference of the SM particles with those NP particles with the same quantum numbers.  Figure~\ref{interference} displays the fractional contribution of the SM+NP interference to the total four-top cross-section, in the parameter space for each of the NP models considered. For many of the relevant benchmarks defined below (see Sect.~\ref{sec:np_ref_points}), the inclusion of electroweak diagrams can account to a modification in the four-top cross-section of up to $\sim 30\%$, all at LO.   

\subsection{NP Benchmark Points}
\label{sec:np_ref_points}
In order to study the NP phenomenology in four-top production, we define in each NP model a set of Benchmark Points (BP) for a representative sample of NP masses and couplings still allowed by the available data.  For each assumed mass ($M$) in a given NP model, we define a tight (T) and a loose (L) BP for which the four-top production cross-section equals 1.5 and 2 times the SM cross-section, respectively.  We denote them as $BP^{NP}_{T,L}(M)$.  Since the spin-0 NP models are excluded by $\gamma\gamma$ resonance searches for masses above 65 GeV we choose the BPs below this mass value.  For other NP models we consider masses of 50 GeV, 150 GeV, and 300 GeV where possible. In Table \ref{rps} we display the values for the couplings and masses in each NP model, which define the BPs.

\begin{table}[ht]
\centering
	\begin{tabular}[t]{ |l|l|l|  }
\multicolumn{2}{c}{Spin-0} \\
\hline
		Benchmark Point &  $g_{\phi/A\,t}$  \\
\hline\hline
$BP^\phi_T (30 \mbox{ GeV})$  & 0.90 \\ 
$BP^\phi_L (30 \mbox{ GeV})$  & 1.09 \\ 
$BP^\phi_T (50 \mbox{ GeV})$  & 0.91 \\ 
$BP^\phi_L (50 \mbox{ GeV})$  & 1.10 \\ 
\hline
$BP^A_T(30 \mbox{ GeV})$ &  1.06\\
$BP^A_L(30 \mbox{ GeV})$ &  1.06\\
$BP^A_T(50 \mbox{ GeV})$ &  1.19\\
$BP^A_L(50 \mbox{ GeV})$ &  1.20\\
\hline
\end{tabular}
	\begin{tabular}[t]{ |l|l|l|  }		
\multicolumn{2}{c}{Spin-1} \\
\hline
		Benchmark Point &  $g_{Z't}$  \\
\hline\hline
$BP^{Z'}_T (50 \mbox{ GeV})$  & 0.20 \\ 
$BP^{Z'}_L (50 \mbox{ GeV})$  & 0.24\\ 
$BP^{Z'}_T (150 \mbox{ GeV})$  & 0.51\\ 
$BP^{Z'}_L (150 \mbox{ GeV})$  & 0.64\\ 
$BP^{Z'}_T(300 \mbox{ GeV})$ &  0.80\\
$BP^{Z'}_L(300 \mbox{ GeV})$ &  0.97\\

\hline
\end{tabular}
	\begin{tabular}[t]{ |l|l|  }
		\multicolumn{2}{c}{Spin-2} \\
\hline
		Benchmark Point & $g_{Gt}$   \\
\hline\hline
		$BP^{G}_T$ (150 GeV)  & 0.33 \\ 
		$BP^{G}_L$ (150 GeV)   & 1.25 \\ 
		$BP^{G}_T$ (300 GeV)   & 0.40 \\ 
		$BP^{G}_L$ (300 GeV)   & 1.49 \\ 
\hline
\end{tabular}
\caption{Benchmark Points selected to study NP effects in four-top phenomenology.  Subscripts $T$ and $L$ stand for tight and loose, for which the four-top production cross-section is 1.5 and 2 times larger than the SM cross-section, respectively. }
\label{rps}
\end{table}

\section{Phenomenology of non-resonant light NP in four-top production}
\label{section:3}

The four-top final state at the LHC represents an exciting opportunity to search for light particles that couple preferentially to the top quark.  In this section we highlight several features in four-top production that are sensitive to this kind of NP contributions.

After decay of the top quarks, a four-top event features a very busy final state with at least 12 energetic partons, including eventual neutrinos.
Therefore, it is extremely challenging the kinematic reconstruction of the final state. In the case of the highest sensitivity channels, 2LSS and ML, the presence of multiple neutrinos makes very difficult the kinematic reconstruction of the leptonically decaying top quarks, although the hadronically decaying top quarks can potentially be reconstructed, particularly if they have significant boost. 

For definiteness, in the following we will restrict our study to the 2LSS channel, which features two same-sign leptons, significant $\met$ because of the presence of two neutrinos, and at least eight jets, four of which are $b$-jets. This choice is appropriate, since the 2LSS channel is one of the most sensitive search channels, although most of our findings will also be applicable to the ML channel, which is dominated by events with exactly three leptons.
We consider several inclusive observables, assuming that kinematic reconstruction is either not available, or too inefficient to be helpful.
One of such observables is the total transverse energy $\hthad$, defined as the scalar sum of the transverse momentum of all jets, leptons and missing energy in the event.  This observable is an trivial extension of the $\hthad^\textrm{jets}$ variable, which only the the jets in the sum, which is typically used by the ATLAS and CMS collaborations in their four-top analyses.\footnote{Note that ATLAS and CMS use ``$\hthad$'' to refer to what we define as $\hthad^\textrm{jets}$.} However we find that $\hthad$, including more information on the event final state objects, is slightly more sensitive.  In any case, we have verified that the results are qualitatively similar when using either of the two variables.  We also define and investigate $tt$ (or $\bar{t}\bar{t}$) spin correlations using the angular separation between the same-sign leptons.  We study the feasibility of using these observables to distinguish NP contributions from the SM, and to discriminate among different NP scenarios.

For this study, we generate $pp \to t \bar t t \bar t$ at LO, including all SM and NP diagrams, for each of the BPs defined in Sect.~\ref{sec:np_ref_points}. The generated events properly account for the helicity transmission in the decay of the two same-sign top quarks, and are showered and processed through a simplified simulation of the ATLAS detector, followed by the reconstruction of detector-level physics objects (see App.~\ref{appendix} for details).
The simulated events are then preselected using requirements based in Ref.~\cite{Sirunyan:2017roi}, which can be summarized as: exactly two same-sign leptons, $\hthad^\textrm{jet}>300$~GeV, $\met > 50$~GeV, and either $\geq 5$ jets of which at least three are $b$-tagged, or $\geq 6$ jets of which at least two are $b$-tagged. 

\subsection{Total transverse energy}
\label{ht}

The total transverse energy $\hthad$ is a variable that provides a measure on how hard the event is, and usually a lower cut in $\hthad$ is used in searches for very massive final states such as those from four-top production, since it suppresses important backgrounds such as $t\bar t$.  The aim of studying this variable in the context of light NP contributions in four-top production is not only to distinguish NP signatures, but also to explore whether $\hthad$ cuts guided by SM four-top searches may inadvertently suppress NP contributions.

\begin{figure}[t]
\begin{center}
\subfloat[]{\includegraphics[width=0.40\textwidth]{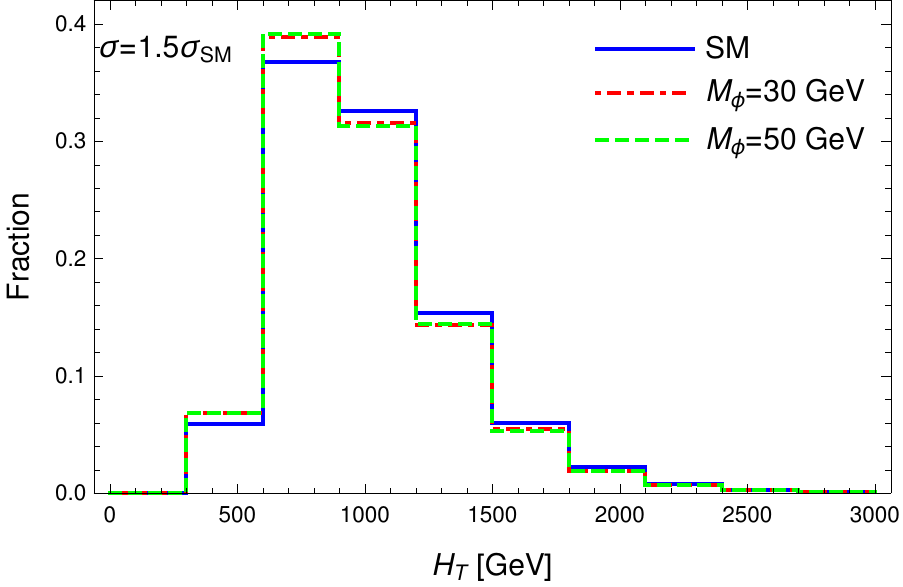}}\hspace{3mm}
\subfloat[]{\includegraphics[width=0.40\textwidth]{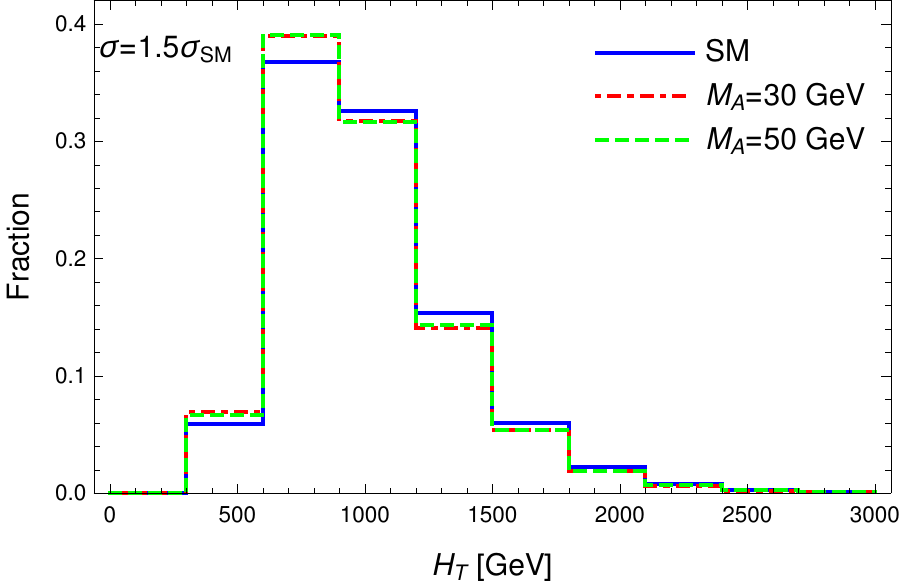}}\\
\subfloat[]{\includegraphics[width=0.40\textwidth]{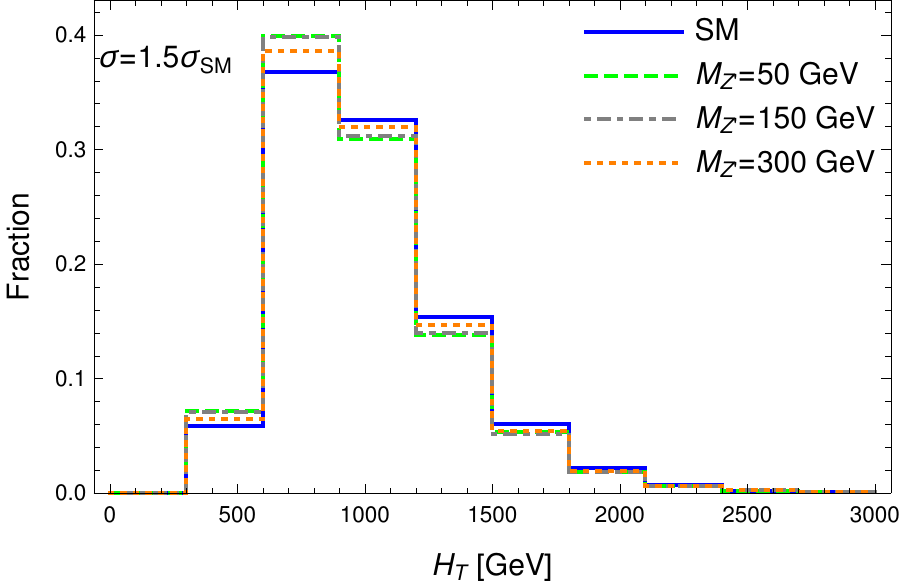}}\hspace{3mm}
\subfloat[]{\includegraphics[width=0.40\textwidth]{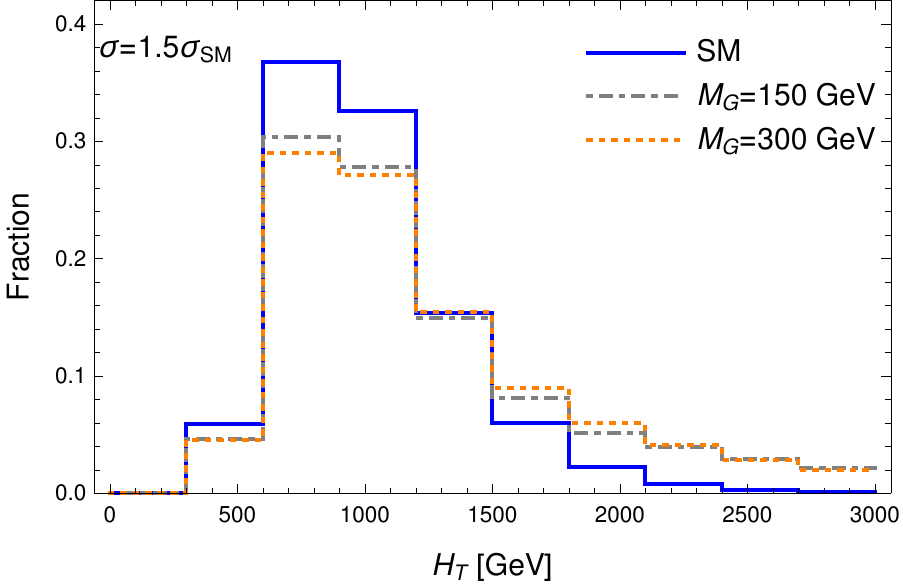}}
\caption{\small Distribution of the $\hthad$ variable for the 2LSS channel after preselection (see text for details), normalized to unit area. Shown are the predictions for the SM (blue) and the tight benchmark points (i.e. giving $\sigma=1.5\sigma_{SM}$) of the NP models considered: (a) Scalar, (b) Pseudo-scalar, (c) $Z'$, and (d) Graviton.  The distinctive behavior in the Graviton is due to the symmetry in one of its main Feynman diagram, as explained in the text. Analogous results for the loose benchmark points (i.e. giving $\sigma=2\sigma_{SM}$) are shown in Fig.~\ref{htscalar2} in App.~\ref{complementaryplots}. }
\label{htscalar}
\end{center}
\end{figure}

Figure~\ref{htscalar} displays the $\hthad$ distribution for each of the NP models considered. Interestingly, in the case of  the spin-0 and spin-1 NP models, the $\hthad$ distribution for SM+NP is found to be slightly softer than that expected for the SM only.  This kind of deviation in the $\hthad$ would typically be attributed to a background mismodeling, and thus potentially missed by current experimental searches. In contrast, the SM+NP distribution for the spin-2 NP model is distinctly harder than the SM prediction, possibly more in line with what is typically expected for ultra-heavy NP, described via an EFT, although in this case we are considering a very light particle.
This can be attributed to the presence of diagrams such as in Fig.~\ref{feynman}c, where there is a symmetry such that the available energy is in average equally distributed among the top quarks in the center-of-mass frame.  Using Lagrange multipliers it can be shown that the maximization of the scalar sum of the top-quark 3-momenta, while constrained to be all on-shell, is obtained for equally distributed energies. This this an effect that increases with the mass/energy ratio of the top quarks.

\subsection{Spin correlations in four-top events}
\label{spincorrelation}
A second observable that is interesting to study in four-top events, and that does not require to reconstruct the four-top system, is the spin correlation between a pair of top quarks. It is easy to appreciate that, depending on the type of the NP particle $X$ exchanged in Fig.\ref{feynman}a, the $ttX$ vertex has  a different Lorentz structure, which in turn affects the spin correlation between the top and antitop quarks in the same fermionic line.  This effect is also transmitted to the same-sign top quarks in different fermionic lines. In this section we study the spin correlation between same-sign top quarks via their corresponding reconstructed leptons in the 2LSS channel.

We first investigate the spin correlation at the parton level by constructing an asymmetry between like and unlike same-sign top-quark helicities. Then, we select events in the 2LSS channel, and study the azimuthal separation in the laboratory frame between the reconstructed same-sign leptons, $\Delta\phi(\ell^\pm,\ell^\pm)$.  In the following paragraphs, we present results for same-sign top quarks ($tt$) and for same-sign positive leptons ($\ell^+\ell^+$), but the same conclusions apply to same-sign antitop quarks ($\bar t \bar t$) and same-sign negative leptons ($\ell^-\ell^-$).

At the parton level one can quantify the spin correlation between the top quarks by defining an asymmetry between the cross-sections for Like ($L$) and Unlike ($U$) top-quark helicities, as given by:
\begin{eqnarray}
	\Att &=& \frac{\sigma(t_+t_+)+\sigma(t_-t_-)-\sigma(t_+t_-)-\sigma(t_-t_+)}{\sigma(t_+t_+)+\sigma(t_-t_-)+\sigma(t_+t_-)+\sigma(t_-t_+)} \,
	\label{atop}
\end{eqnarray}
\noindent where $\sigma(t_i t_j)$ denotes the cross-section for $t\bar t t \bar t$ production with the two top quarks ($tt$) having helicities $i$ and $j$, respectively, summed over the antitop-quark helicities. 

\begin{figure}[t!]
\begin{center}
\subfloat[]{\includegraphics[width=0.40\textwidth]{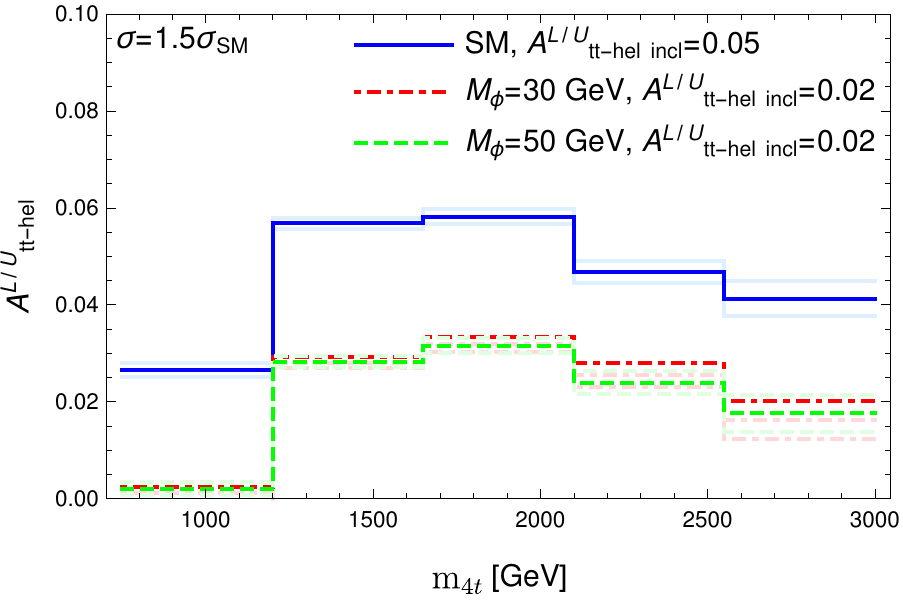}}\hspace{3mm}
\subfloat[]{\includegraphics[width=0.40\textwidth]{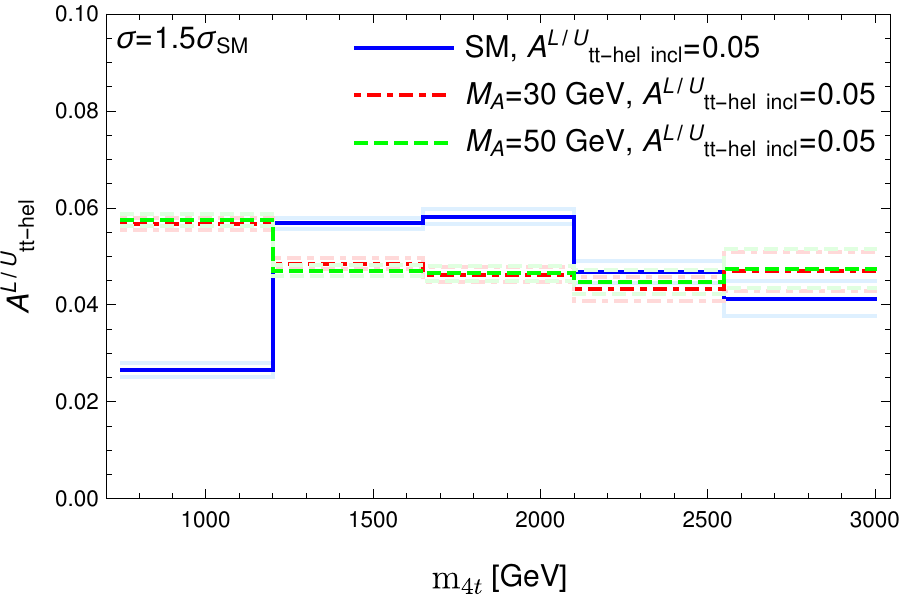}}\\
\subfloat[]{\includegraphics[width=0.40\textwidth]{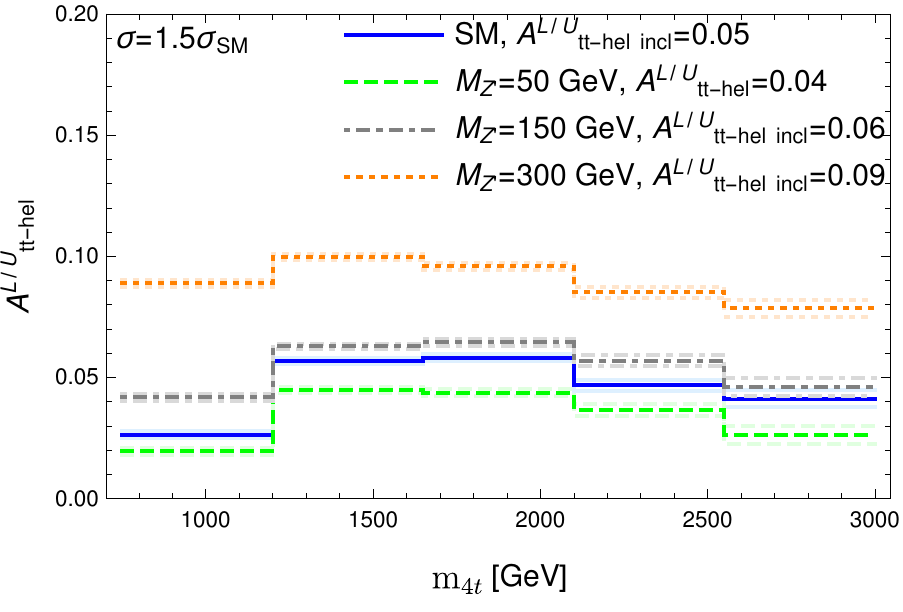}}\hspace{3mm}
\subfloat[]{\includegraphics[width=0.40\textwidth]{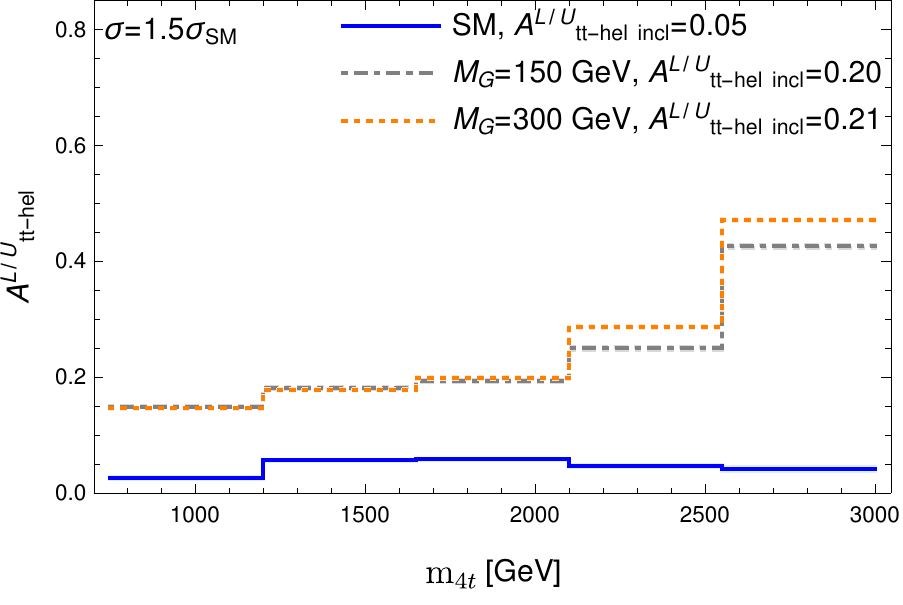}}
\caption{\small $\Att$ asymmetry (see Eq.~\ref{atop}) as a function of the invariant mass of the four-top system. Shown are the predictions for the SM (blue) and the tight benchmark points (i.e. giving $\sigma=1.5\sigma_{SM}$) of the NP models considered: (a) Scalar, (b) Pseudo-scalar, (c) $Z'$, and (d) Graviton.  Also quoted are the inclusive asymmetries, i.e. averaged over the four-top invariant mass spectrum. Analogous results for the loose benchmark points (i.e. giving $\sigma=2\sigma_{SM}$) are shown in Fig.~\ref{Att2} in App.~\ref{complementaryplots}.}
\label{Att}
\end{center}
\end{figure}

Figure~\ref{Att} displays the $\Att$ asymmetry as a function of the four-top invariant mass in the SM, as well as for the different NP BPs considered.  We find negative contributions to $\Att$ in the Scalar case, a slight negative contribution in the Pseudo-scalar case, a positive (negative) contribution in the $Z'$ case with high (low) mass, and a large positive contribution in the Graviton case.  The two latter results could be expected, since the $Z'$ and Graviton contributions include only right-chiral top quarks which, at higher energy, are likely to have positive helicity. 

\begin{figure}[t!]
\begin{center}
\subfloat[]{\includegraphics[width=0.40\textwidth]{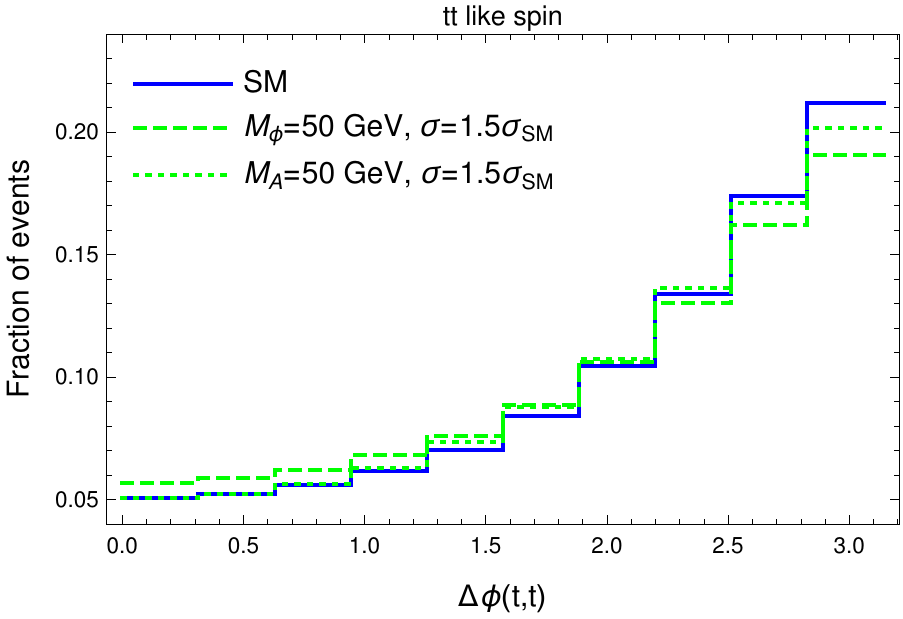}}\hspace{3mm}
\subfloat[]{\includegraphics[width=0.40\textwidth]{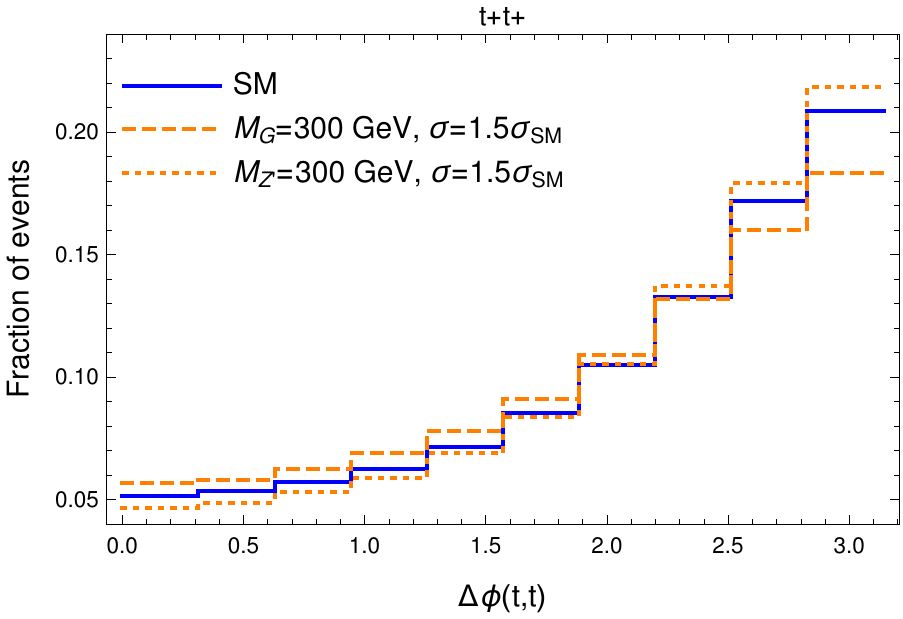}}	
\caption{\small Distribution of the azimuthal separation between the same-sign top quarks, $\Delta\phi(t,t)$, normalized to unit area. Shown are the predictions for the SM (blue) and the tight benchmark points (i.e. giving $\sigma=1.5\sigma_{SM}$) of the NP models considered for the dominant helicity configurations: (a) Scalar and Pseudo-scalar, (b) $Z'$ and Graviton.}
\label{tt-distribution}
\end{center}
\end{figure}

In contrast to $t\bar t$ production, since $t\bar t t \bar t$ is a four-body final state, the translation from these top-quark polarization asymmetries to the angular separation between leptons is not straightforward.  In fact, the angular separation between the top quarks, which depends on the underlying dynamics, also affects the angular separation between the final-state leptons.  
This distribution, together with $A^{L/U}_\textrm{tt-hel}$ (see Fig.~\ref{Att}), provide some insights of what can be expected for the angular distribution of the top-quark-decay products.
In Fig.~\ref{tt-distribution} we display the azimuthal separation between the two same-sign top quarks, $\Delta\phi(t,t)$, for the dominant helicity configurations in the each of the NP scenarios considered.
In particular, for the $Z'$ and Graviton modes we show the $t_+ t_+$ configuration because this final state represents 33\% and 47\% of the total cross-section, respectively.  A suppression (enhancement) in the back-to-back configurations for the top quarks in these helicity configurations tends to suppress (enhance) the back-to-back configuration between their corresponding same-sign leptons.

\begin{figure}[t!]
\begin{center}
\subfloat[]{\includegraphics[width=0.40\textwidth]{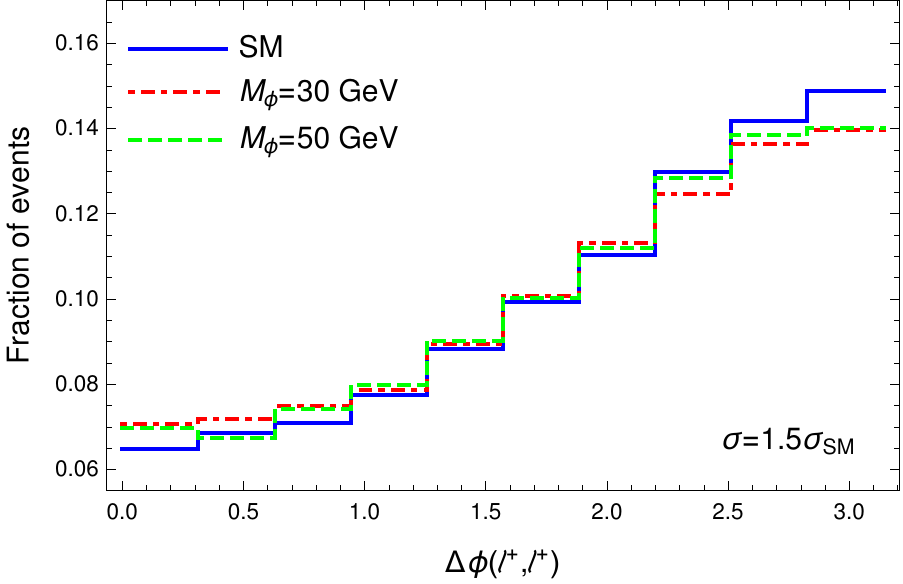}}\hspace{3mm}
\subfloat[]{\includegraphics[width=0.40\textwidth]{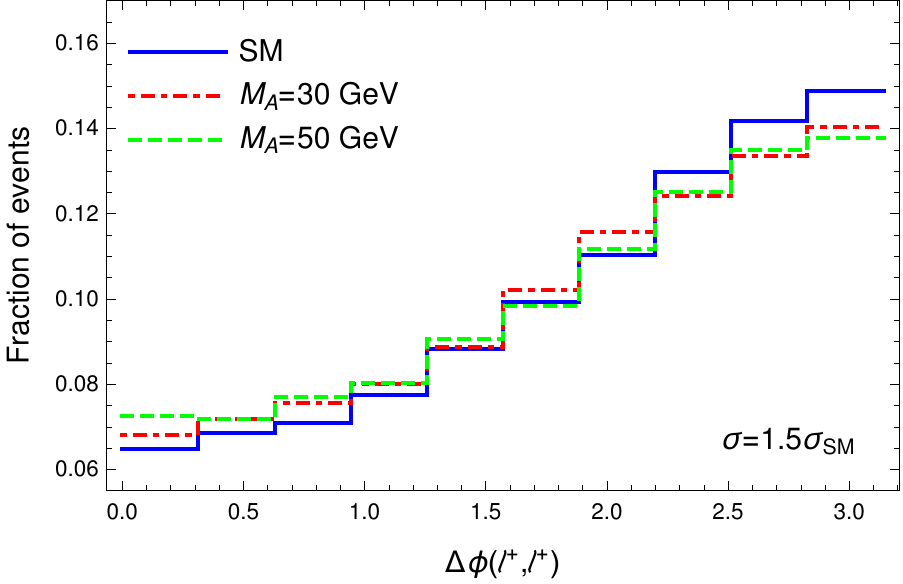}}\\
\subfloat[]{\includegraphics[width=0.40\textwidth]{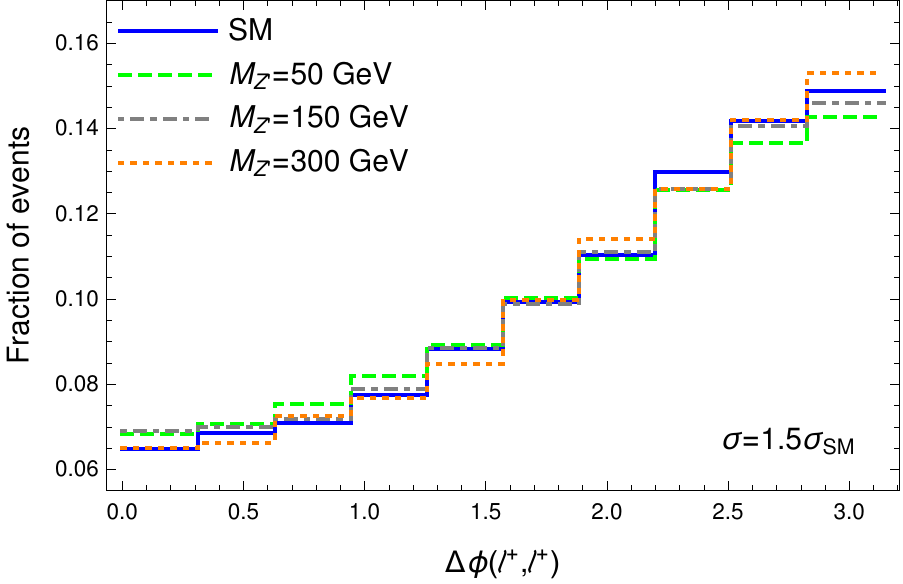}}\hspace{3mm}
\subfloat[]{\includegraphics[width=0.40\textwidth]{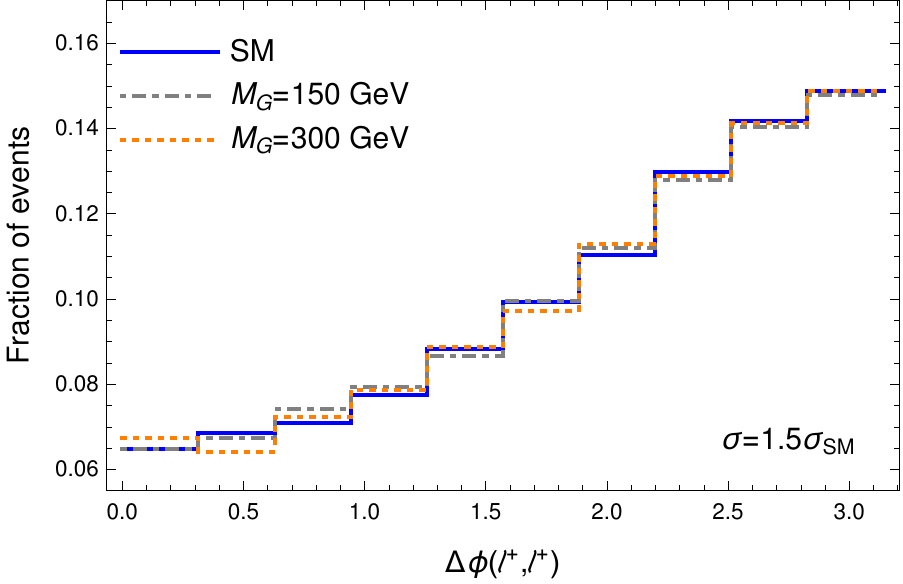}}
\caption{\small Distribution of the azimuthal separation between the same-sign leptons,  $\Delta\phi(\ell^+,\ell^+)$, for the 2LSS channel after preselection (see text for details), normalized to unit area. Shown are the predictions for the SM (blue) and the tight benchmark points (i.e. giving $\sigma=1.5\sigma_{SM}$) of the NP models considered: (a) Scalar, (b) Pseudo-scalar, (c) $Z'$, and (d) Graviton.  Analogous results for the loose benchmark points (i.e. giving $\sigma=2\sigma_{SM}$) are shown in Fig.~\ref{dphi-inclusive2} in App.~\ref{complementaryplots}. 
These same results in two bins of $\hthad$ are shown in Figs.~\ref{dphiscalar}--\ref{dphigraviton} in App.~\ref{complementaryplots}.}
\label{dphi-inclusive}
\end{center}
\end{figure}

When considering spin-correlation observables in four-top events, the simplest observable is the azimuthal separation between same-sign leptons, $\Delta\phi(\ell^+,\ell^+)$, in the 2LSS and ML channels.  
Figure~\ref{dphi-inclusive} shows a comparison of the predicted $\Delta\phi(\ell^+,\ell^+)$ distribution in the 2LSS channel after preselection, between the SM and the tight benchmark points (i.e. giving $\sigma=1.5\sigma_{SM}$) of the NP models considered.
As can be appreciated, the Scalar and Pseudo-scalar models are characterized by a depletion of back-to-back SS leptons compared to the SM.  In the case of $Z'$ model, there is an enhancement (depletion) of back-to-back SS leptons for high (low) mass. Surprisingly, the Graviton model does not display a significant difference with respect to the SM distribution.
This appears to be (at least partly) explained by an accidental cancellation of effects in the parton-level polarizations (see Fig.~\ref{Att}) and the parton-level angular distributions (see Fig.~\ref{tt-distribution}).  

\section{Discussion}
\label{section:4}

The four-top signal at the LHC is a relatively new subject and the community is in the course of acquiring and processing knowledge on the many aspects of this final state.  We have presented a set of results that raise new questions and challenges, which we discuss in the following paragraphs.  We begin with a discussion concerning parton-level four-top production and then we examine the results concerning final state particles and detector-level results.

In this work we have simulated four-top production at LO, and we have applied a $k$-factor to estimate NLO corrections to the total cross section.  However, we have found that for the studied benchmark points at LO, the interference between SM QCD+QED and NP amplitudes can account for a large fraction of the total cross-section, $\sim 30\%$.  This suggests that a simulation of four-top production at NLO, including NP contributions, is required for more precise interpretation of experimental results in terms of the parameter space of the NP models considered.  This conclusion had been envisaged in previous NLO studies of SM four-top production~\cite{Frederix:2017wme,Bevilacqua:2012em}.

Since the reconstruction of the four-top final state is very challenging, one of the interesting observables after event selection is the distribution of the total transverse energy, $\hthad$.  The presented NP scenarios with light resonances that cannot be produced on-shell, tend to produce an $\hthad$ spectrum softer than the SM in the cases of spin-0 and spin-1 NP, whereas the spectrum is harder in the case of spin-2 NP.  Interestingly, the latter is analogous to the effect from a heavy off-shell resonance \cite{Degrande:2010kt}. The spin-2 NP model differs from the other models in the presence of extra Feynman diagrams that involve four-point interactions (see Fig.~\ref{feynman}c).  A softer $\hthad$ spectrum for spin-0 and spin-1 NP could lead to a bias in the measured four-top cross-section \cite{Sirunyan:2019wxt}, since SM kinematics is usually assumed to estimate the acceptance and shape of the final discriminating variable. Further work in this direction would be interesting.  Nevertheless, our results suggest that $\hthad$ is a useful observable to discriminate possible NP contributions, even in the case of new light particles.

The combined study of the parton-level polarization asymmetries and angular distributions, e.g. $\Delta\phi(\ell^+,\ell^+)$ in the 2LSS channel, is suggestive of a rich and exciting program yet to be developed related to the use of this kind of observables. Whereas the $\Delta\phi(\ell^+\ell^+)$ distribution can be used to probe spin-0 NP scenarios, the Graviton and $Z'$ models would be better probed by studying the angular separation between two reconstructed top quarks. This means that new opportunities arise if one could reconstruct the top quarks.  This could be achieved with large statistics in the 2LSS channel, using some of the sophisticated reconstruction algorithms already in use by the experimental collaborations. Alternatively, in the ML channel, useful information could also be extracted from the angular separation between one lepton and the reconstructed top quark. In general, the study of polarization effects in four-top production is an attractive field that requires further investigation.

Our study has focused on the spin correlation between same-sign top quarks in the 2LSS channel.  We have shown that this observable is sensitive to NP,  even though these same-sign top quarks do not share a common vertex in the Feynman diagrams.  We consider that a similar analysis, but in the opposite-sign dilepton channel  --i.e. with leptons coming from opposite-sign top quarks--, could be potentially interesting. This has the advantage that opposite-sign top quarks can share a vertex in the Feynman diagrams, and therefore their relative spin would be sensitive to the Lorentz structure of the underlying physics. On the other hand, the 2LOS channel has significantly lower signal-to-background ratio than the 2LSS and ML channels, plus in half of the cases the opposite-sign top quarks would not share a common vertex in the Feynman diagrams, thus potentially affecting the sensitivity.

The observables studied in this work could be helpful towards establishing an eventual deviation in four-top production. The level of model discrimination of these observables indicates that they could be exploited by the experimental analyses using LHC Run 3 data and beyond.  In any case, the smoking gun for a light new particle with spin-0 ($H$ or $A$) or spin-2 ($G$) could come from $pp \to H/A/G \to \gamma\gamma$ resonant production searches. Even more promising could be the study of the di-photon invariant mass spectrum  in $pp\to t\bar t H/A/G (\to \gamma\gamma)$ production, owing to the more favourable signal-to-background ratio.  We note that the $t\bar t \gamma \gamma$ final state has been studied so far only for a resonance in the SM Higgs mass region, and thus the extension of this search to a broader mass range would be extremely interesting. In the case of a light resonance with spin-1 (e.g. a $Z'$), potentially interesting processes would be $pp\to Z'j$ and $pp \to t \bar t Z'$, with $Z' \to \gamma\gamma^*\to \gamma \ell^+\ell^-$. Observe that the one-loop Feynman diagrams $gg \to Z'g$ and $Z'\to ggg$ with tops running in the internal lines, whose features can be found in Ref.~\cite{vanderBij:1988ac}, are key ingredients to study the previous processes. Further studies in these directions would be interesting as well.

\section{Conclusions}
\label{section:5}

We have studied the phenomenology of four-top production at the LHC for a variety of simple NP models consisting in a top-philic resonance whose mass is below the $t\bar{t}$ threshold.  We have analyzed observables at parton and detector level and studied how they could be used to probe NP contributions, as well as discriminate among them.

The investigated NP models include a light Scalar, Pseudo-scalar, vector $Z'$, and Graviton.  Lorentz invariance in spin-0 resonances requires both top-quark chiralities in the interaction, whereas spin-1 and spin-2 models can be set to couple only to $t_R$, being less constrained by $SU(2)_L$ precision tests.  The Graviton non-renormalizable Lagrangian includes an extra set of four-point interactions --involving two top quarks, a Graviton and a gauge boson-- which provides a distinguishing feature for the model.

We have focused our study in regions of parameter space where the four-top production cross-section is 1.5 and 2 times the SM-expected cross-section, which is consistent with the latest experimental results.  We have found that these regions are very sensitive to tree-level QED corrections when NP contributions are included, indicating that full NLO predictions including NP contributions would be an important development for the correct interpretation of future experimental results.   We have found that available $\gamma\gamma$ resonance searches exclude masses above 65 GeV for the spin-0 models, while the spin-2 model remains largely unconstrained.  
In the remaining allowed parameter space, we have defined some benchmark points and studied a set of observables and their phenomenology.

We have studied the 2LSS channel, which is one of the most sensitive final state signatures being probed experimentally, We have studied the distribution of the scalar sum of all objects $p_T$, $\hthad$, which is widely used by the experimental searches at the LHC.  We have found that, in comparison to the SM, the spin-0 and spin-1 models predict a softer spectrum, whereas the spin-2 model predicts a harder spectrum.  We conclude that such a change in shape towards the softer spectrum region for spin-0 and spin-1 could be translated into an incorrect estimation of the measured four-top production cross-section.  On the other hand, the harder $\hthad$-spectrum in the Graviton model would be a valuable discriminating feature for this model.   

Given the different Lorentz structure of the interactions in each NP model, we have also investigated the spin correlation in four-top production, and its traces in the final-state particles.  At the parton-level, we have studied the relative helicity of both top quark by defining a top-quark Like/Unlike helicity asymmetry (see Eq.~\ref{atop}) and comparing the predictions from the SM and the different NP models considered.  We have found negative contributions to the asymmetry for the spin-0 models and positive contributions for a high-mass $Z'$ and Graviton.  In order to relate this helicity asymmetry to the azimuthal separation between same-sign leptons,  $\Delta\phi(\ell^\pm,\ell^\pm)$, we have also studied the parton-level azimuthal separation between the same-sign top quarks.  We have found that the $\Delta\phi(\ell^\pm,\ell^\pm)$ distribution can be particularly sensitive to spin-0 NP.

We have included a discussion section where we examine the results in the article.  We consider that the available results provide in principle a set of tools that would be useful, not only to detect the presence of light non-resonant NP in four-top production, but also to determine the nature of this NP.  We find that to convert clues from these observables into hard evidence, resonance searches in the $\gamma\gamma$ and $t\bar t \gamma\gamma$ channels would be crucial in all cases, except for the spin-1 NP.  In the latter case, the corroboration could come by resonance searches replacing $\gamma\gamma$ by $\gamma \gamma^* \to \gamma \ell^+\ell^-$.

This article should be considered a first approach in studying the aforementioned observables within the presented simple NP models.  To have a more realistic estimation on the significance of to what extent the available results could probe the NP in four-top production, a more comprehensive analysis including NLO calculations and backgrounds should be performed on the channels and observables as described above.  Nevertheless, the outcome of our work shows that such a study would be very relevant for the upcoming four-top phenomenology.

Four-top studies at experimental, phenomenological,  and theoretical levels are becoming a powerful tool to investigate light top-philic NP.  The community is currently at an stage of learning and developing new tools and features on this interesting final state.  We expect four-top to be an important field in the forthcoming years and for the HL-LHC.

\section*{Acknowledgments}
We thank Leandro Da Rold, Daniel de Florian, Mariel Estévez and Manuel Szewc for useful conversations.

\appendix

\section{Complementary plots}
\label{complementaryplots}
\begin{figure}[h!]
\begin{center}
\subfloat[]{\includegraphics[width=0.40\textwidth]{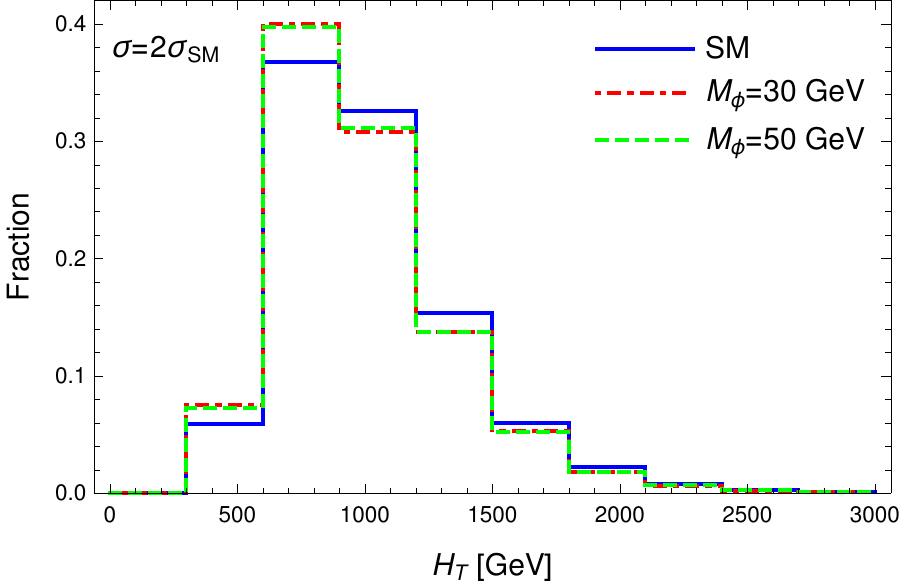}}\hspace{3mm}
\subfloat[]{\includegraphics[width=0.40\textwidth]{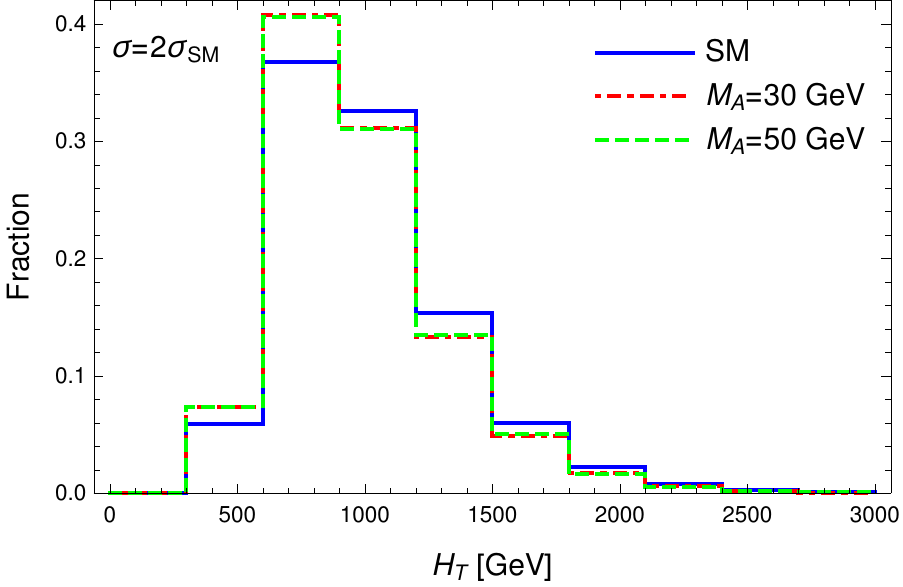}}\\
\subfloat[]{\includegraphics[width=0.40\textwidth]{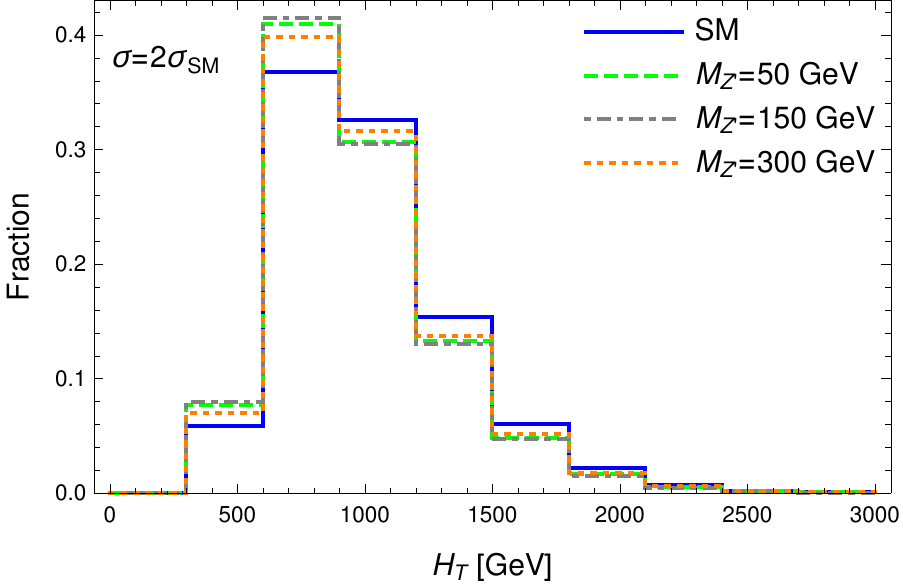}}\hspace{3mm}
\subfloat[]{\includegraphics[width=0.40\textwidth]{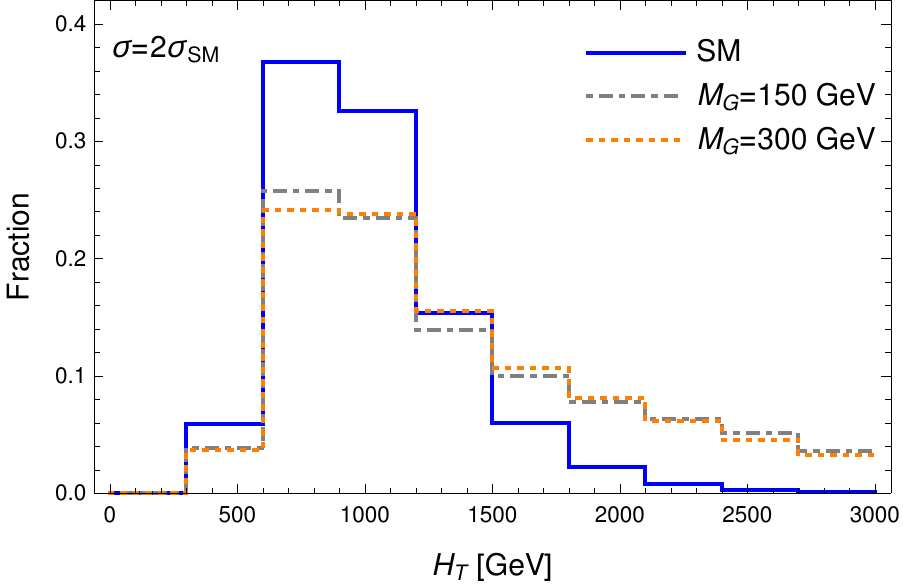}}
 \caption{\small Distribution of the $\hthad$ variable for the 2LSS channel after preselection (see text for details), normalized to unit area. Shown are the predictions for the SM (blue) and the loose benchmark points (i.e. giving $\sigma=2\sigma_{SM}$) of the NP models considered: (a) Scalar, (b) Pseudo-scalar, (c) $Z'$, and (d) Graviton.  
The distinctive behavior in the Graviton is due to the the symmetry in one of its main Feynman diagram, as explained in the text. Analogous results for tight benchmark points are shown in Fig.~\ref{htscalar}.}
\label{htscalar2}
\end{center}
\end{figure}

\begin{figure}[h!]
\begin{center}
\subfloat[]{\includegraphics[width=0.40\textwidth]{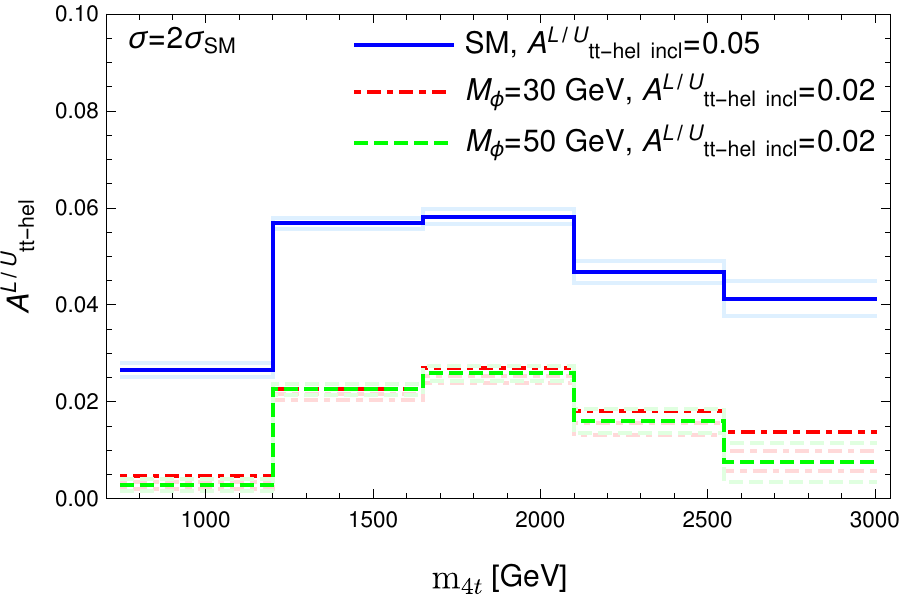}}\hspace{3mm}
\subfloat[]{\includegraphics[width=0.40\textwidth]{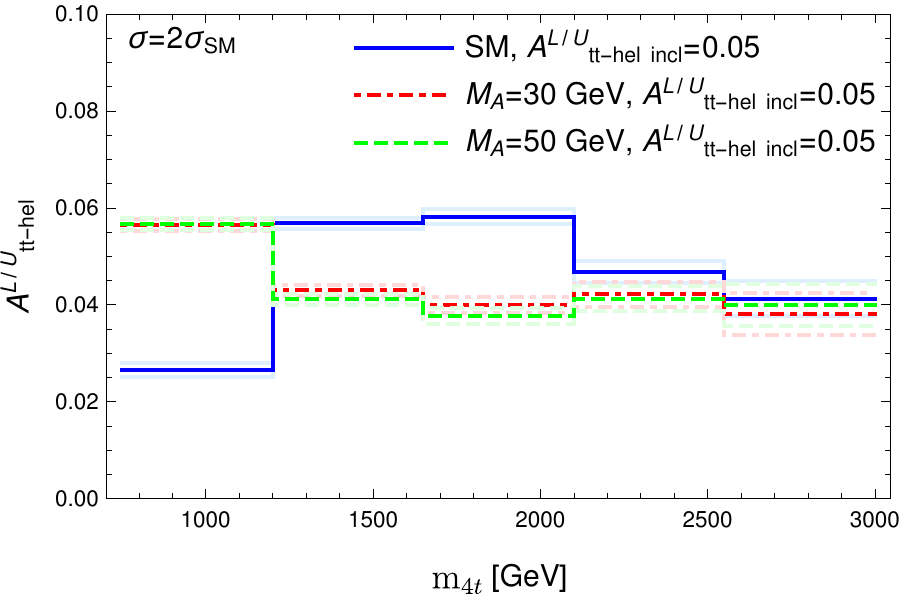}}\\
\subfloat[]{\includegraphics[width=0.40\textwidth]{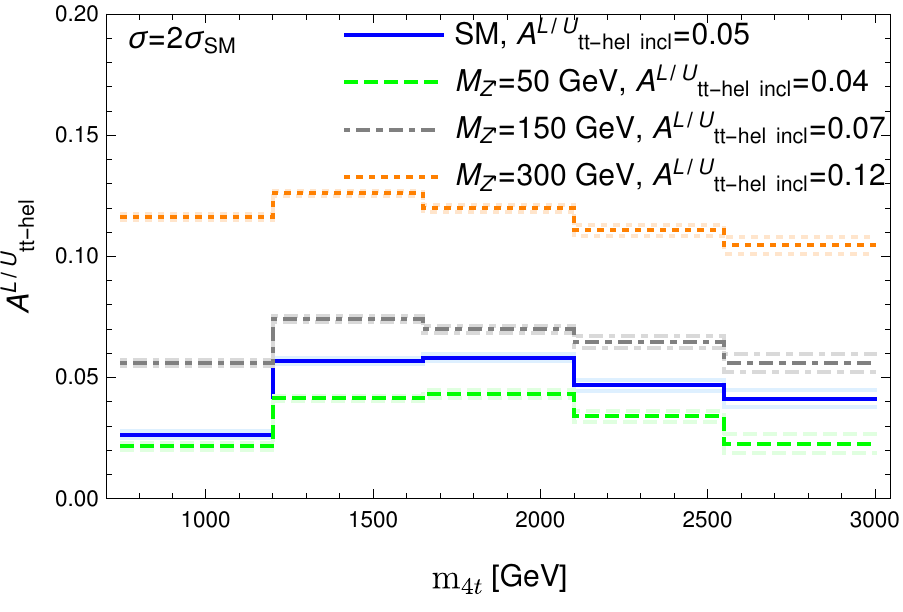}}\hspace{3mm}
\subfloat[]{\includegraphics[width=0.40\textwidth]{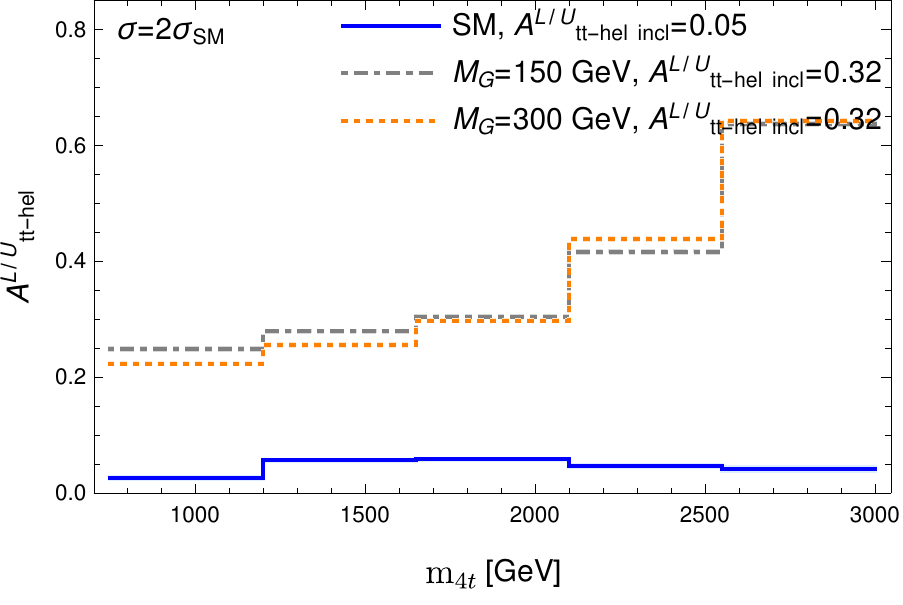}}
\caption{\small $\Att$ asymmetry (see Eq.~\ref{atop}) as a function of the invariant mass of the four-top system. Shown are the predictions for the SM (blue) and the loose benchmark points (i.e. giving $\sigma=2\sigma_{SM}$) of the NP models considered: (a) Scalar, (b) Pseudo-scalar, (c) $Z'$, and (d) Graviton.  Also quoted are the inclusive asymmetries, i.e. averaged over the four-top invariant mass spectrum. Analogous results for the tight benchmark points (i.e. giving $\sigma=1.5\sigma_{SM}$) are shown in Fig.~\ref{Att}.}
\label{Att2}
\end{center}
\end{figure}

\begin{figure}[h!]
\begin{center}
\subfloat[]{\includegraphics[width=0.40\textwidth]{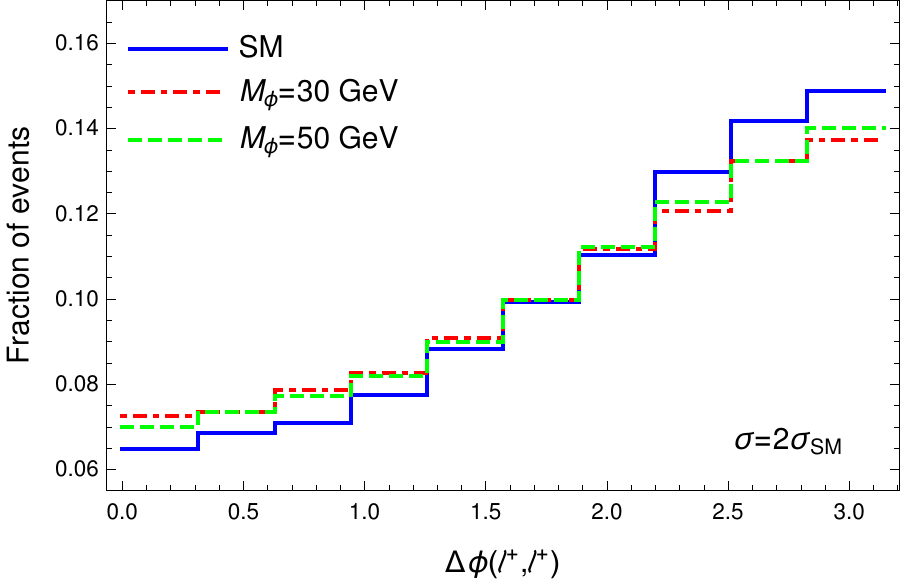}}\hspace{3mm}
\subfloat[]{\includegraphics[width=0.40\textwidth]{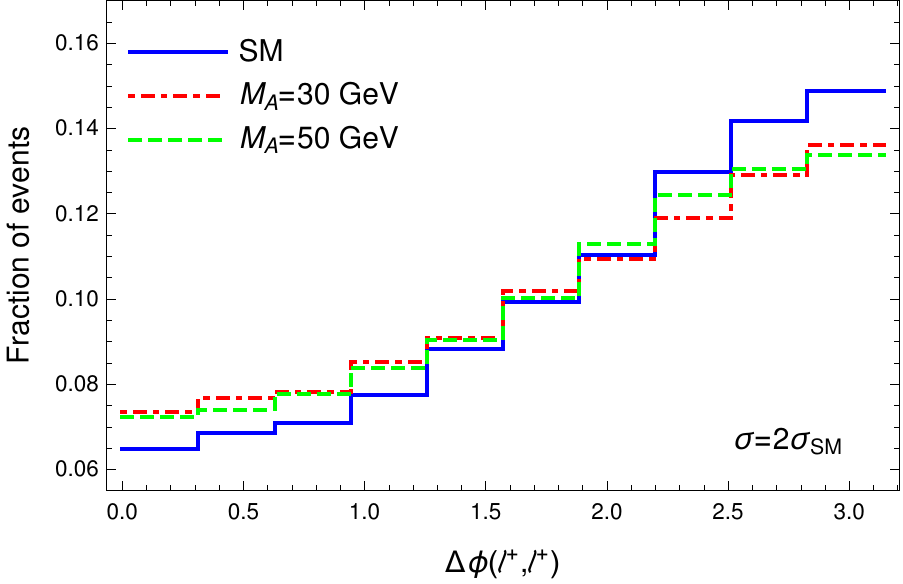}}\\
\subfloat[]{\includegraphics[width=0.40\textwidth]{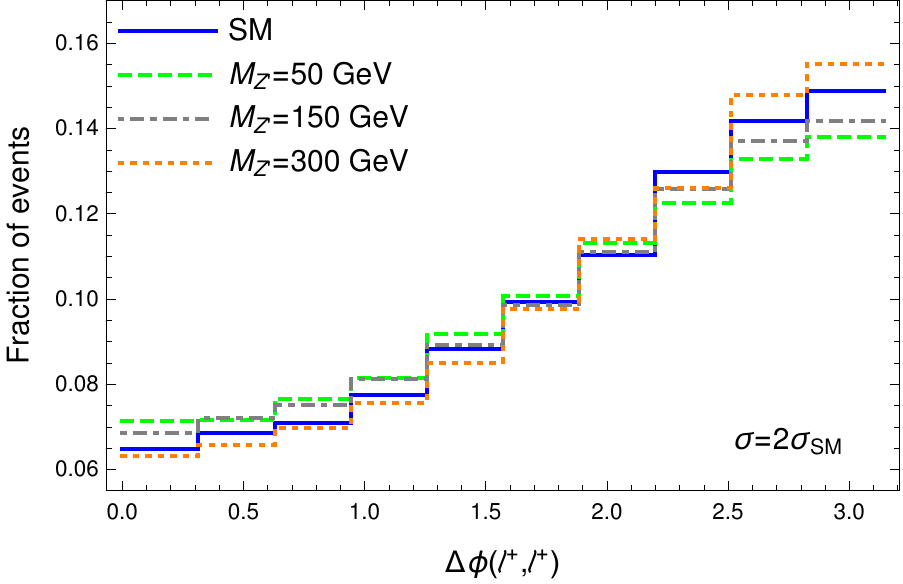}}\hspace{3mm}
\subfloat[]{\includegraphics[width=0.40\textwidth]{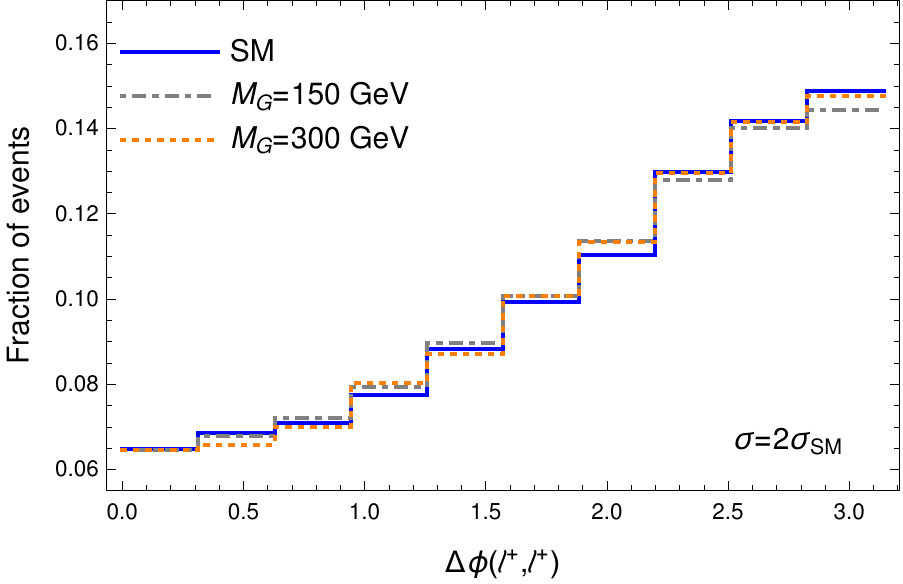}}
\caption{\small Distribution of the azimuthal separation between the same-sign leptons,  $\Delta\phi(\ell^+,\ell^+)$, for the 2LSS channel after preselection (see text for details), normalized to unit area. Shown are the predictions for the SM (blue) and the loose benchmark points (i.e. giving $\sigma=2\sigma_{SM}$) of the NP models considered: (a) Scalar, (b) Pseudo-scalar, (c) $Z'$, and (d) Graviton.  Analogous results for the tight benchmark points (i.e. giving $\sigma=1.5\sigma_{SM}$) are shown in Fig.~\ref{dphi-inclusive}. }
\label{dphi-inclusive2}
\end{center}
\end{figure}

\begin{figure}[h!]
\begin{center}
\subfloat[]{\includegraphics[width=0.40\textwidth]{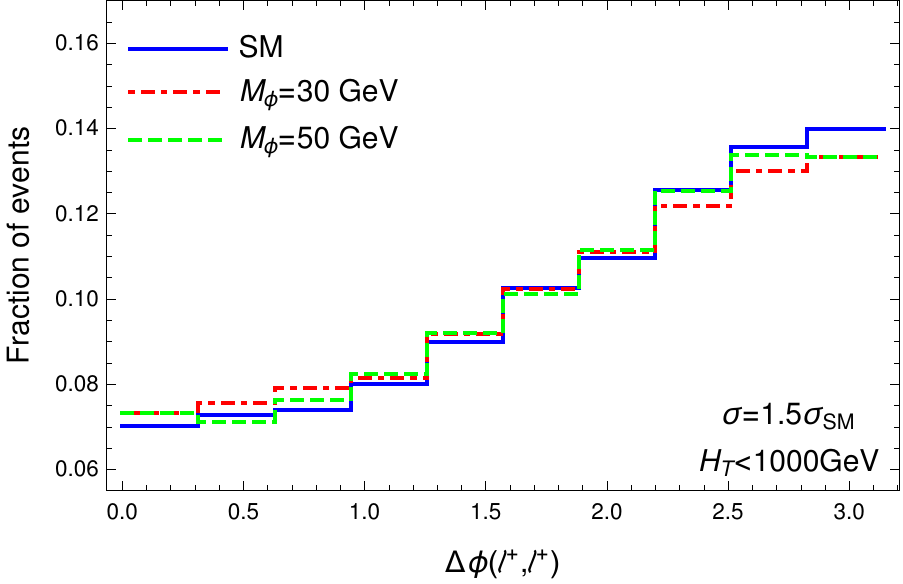}}\hspace{3mm}
\subfloat[]{\includegraphics[width=0.40\textwidth]{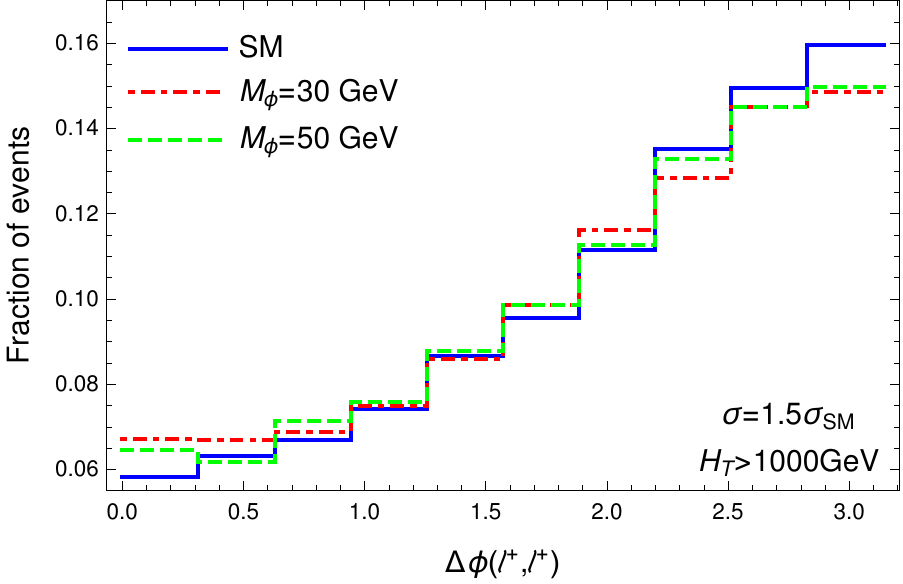}} \\
\subfloat[]{\includegraphics[width=0.40\textwidth]{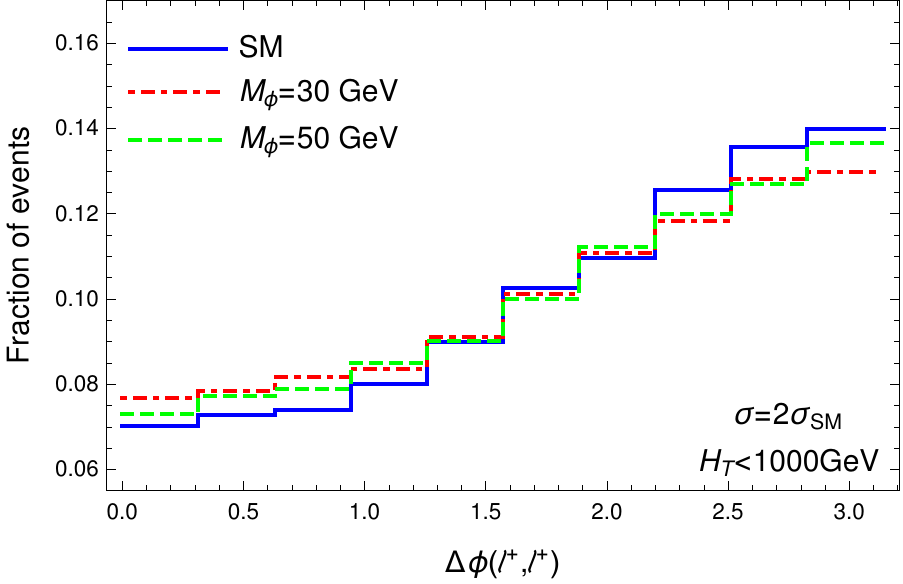}}\hspace{3mm}
\subfloat[]{\includegraphics[width=0.40\textwidth]{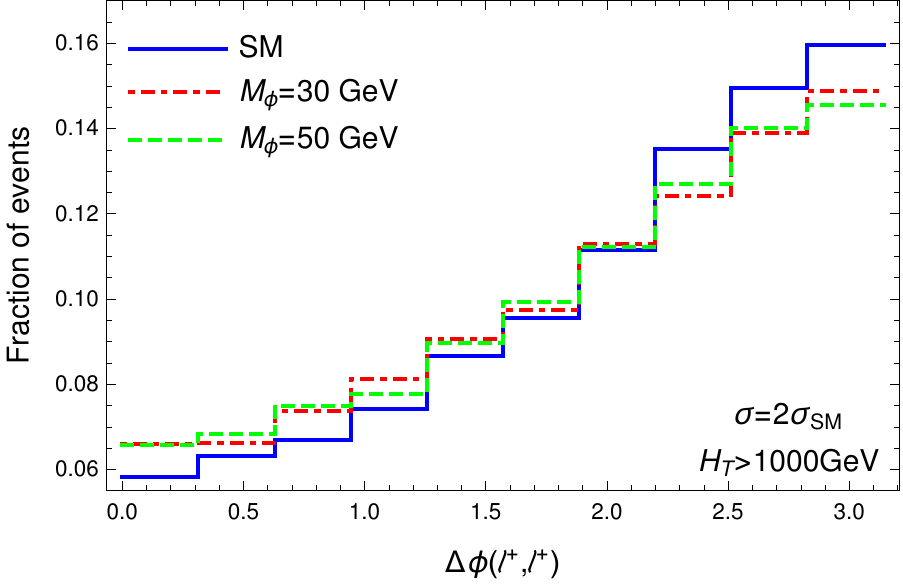}} \\
\caption{\small Distribution of the azimuthal separation between the same-sign leptons,  $\Delta\phi(\ell^+,\ell^+)$, for the 2LSS channel after preselection (see text for details) with 
(left) $\hthad<1000$~GeV and (right) $\hthad>1000$~GeV, normalized to unit area.  
Shown are the predictions for the SM (blue) and the Scalar model for (a, b) the tight benchmark points (i.e. giving $\sigma=1.5\sigma_{SM}$) and
(c, d) the loose benchmark points (i.e. giving $\sigma=2\sigma_{SM}$).}
\label{dphiscalar}
\end{center}
\end{figure}

\begin{figure}[h!]
\begin{center}
\subfloat[]{\includegraphics[width=0.40\textwidth]{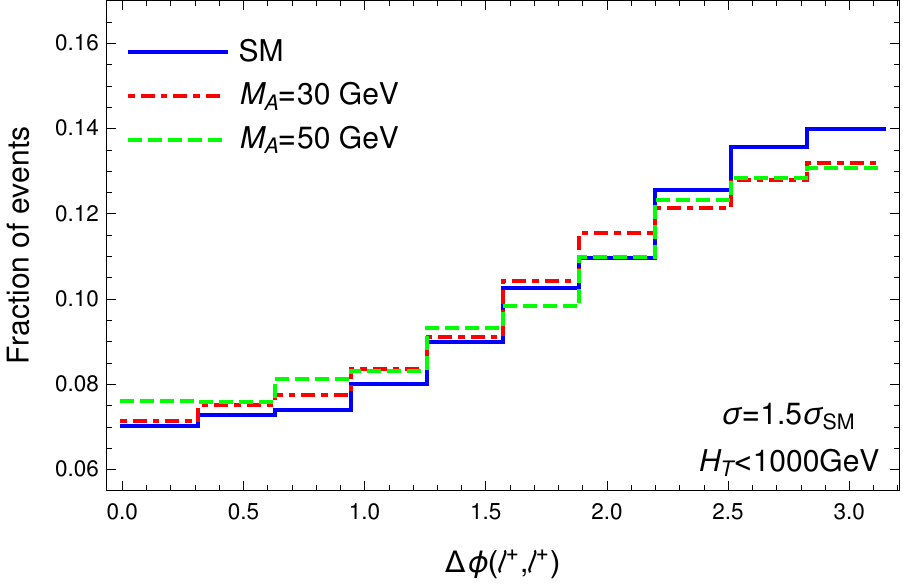}}\hspace{3mm}
\subfloat[]{\includegraphics[width=0.40\textwidth]{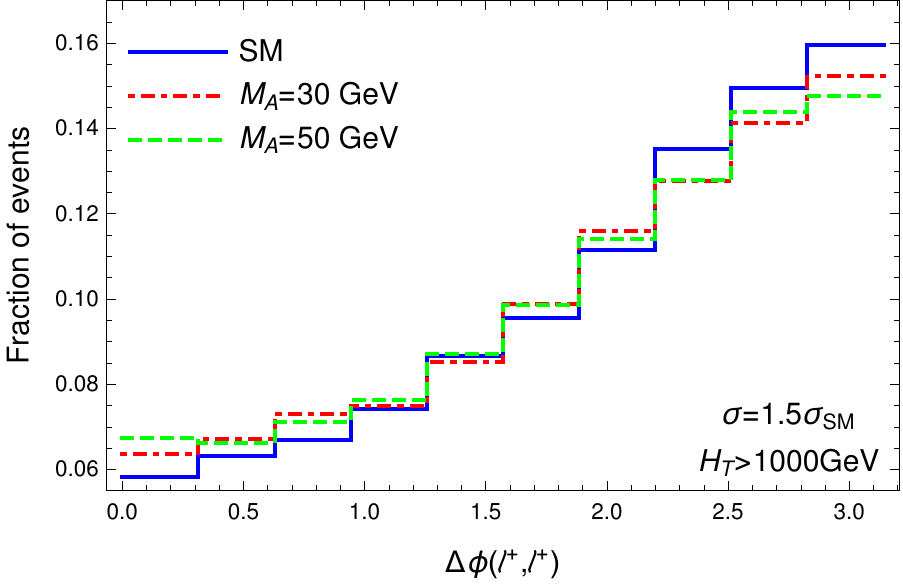}} \\
\subfloat[]{\includegraphics[width=0.40\textwidth]{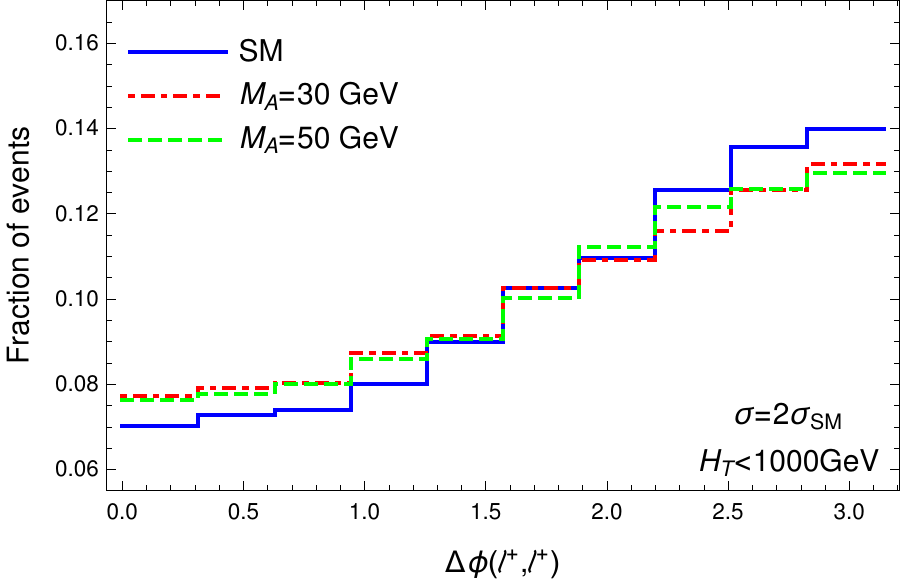}}\hspace{3mm}
\subfloat[]{\includegraphics[width=0.40\textwidth]{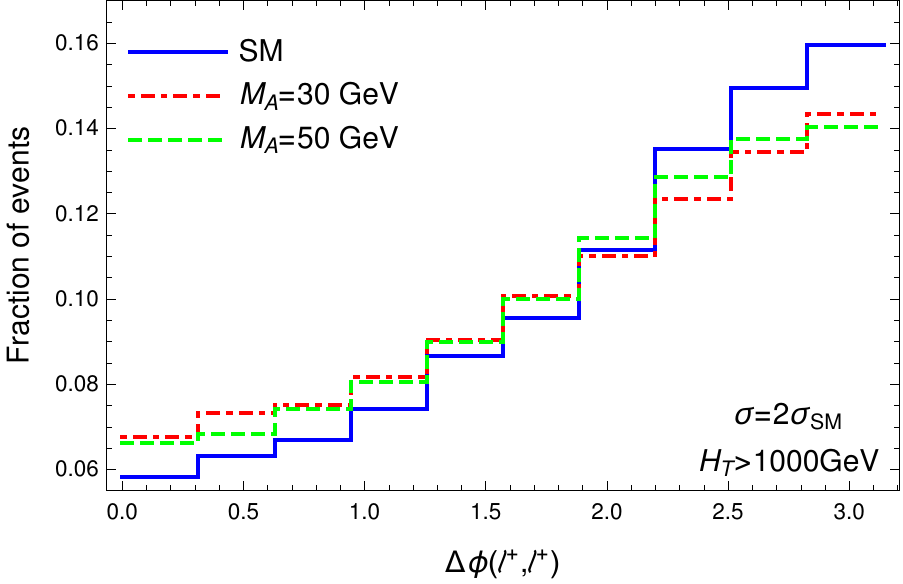}} \\
\caption{\small Distribution of the azimuthal separation between the same-sign leptons,  $\Delta\phi(\ell^+,\ell^+)$, for the 2LSS channel after preselection (see text for details) with 
(left) $\hthad<1000$~GeV and (right) $\hthad>1000$~GeV, normalized to unit area.  
Shown are the predictions for the SM (blue) and the Pseudo-scalar model for (a, b) the tight benchmark points (i.e. giving $\sigma=1.5\sigma_{SM}$) and
(c, d) the loose benchmark points (i.e. giving $\sigma=2\sigma_{SM}$).}
\label{dphipseudoscalar}
\end{center}
\end{figure}

\begin{figure}[h!]
\begin{center}
\subfloat[]{\includegraphics[width=0.40\textwidth]{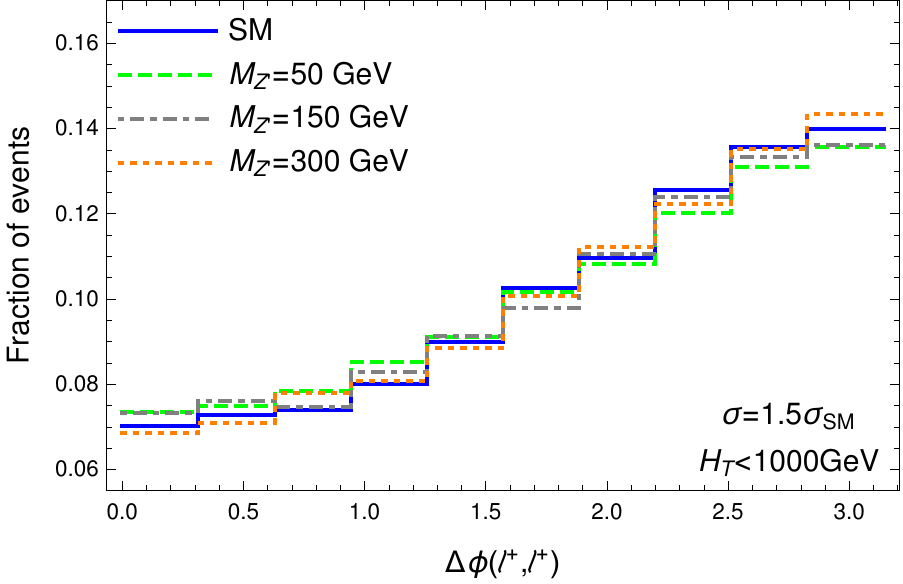}}\hspace{3mm}
\subfloat[]{\includegraphics[width=0.40\textwidth]{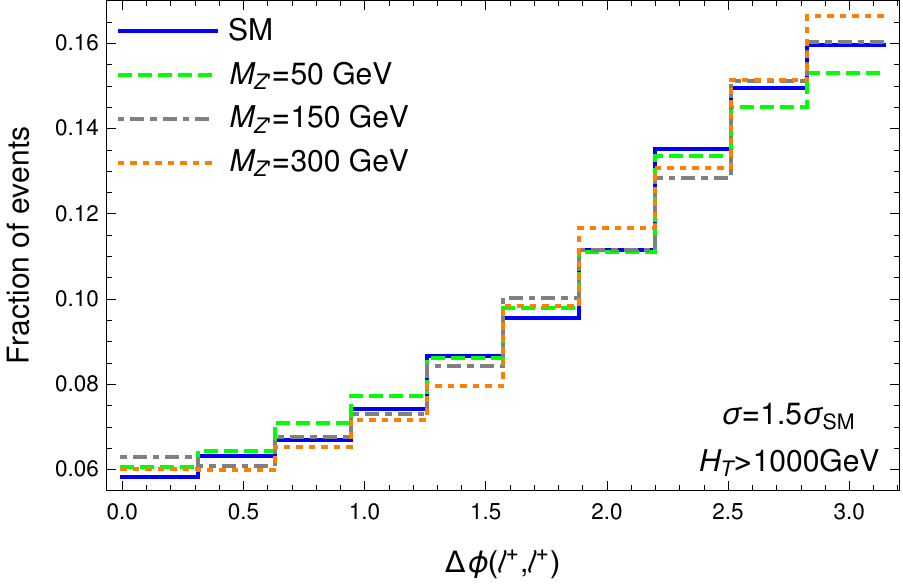}} \\
\subfloat[]{\includegraphics[width=0.40\textwidth]{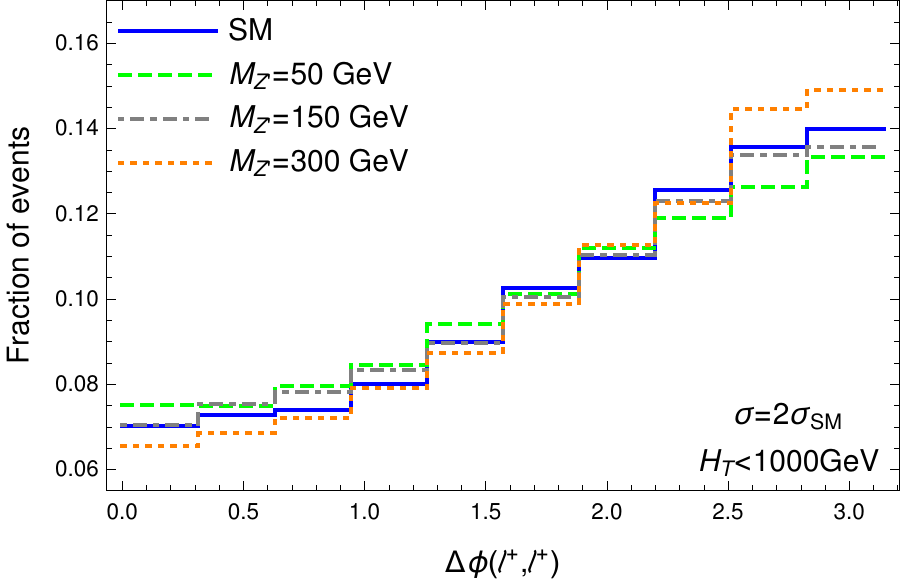}}\hspace{3mm}
\subfloat[]{\includegraphics[width=0.40\textwidth]{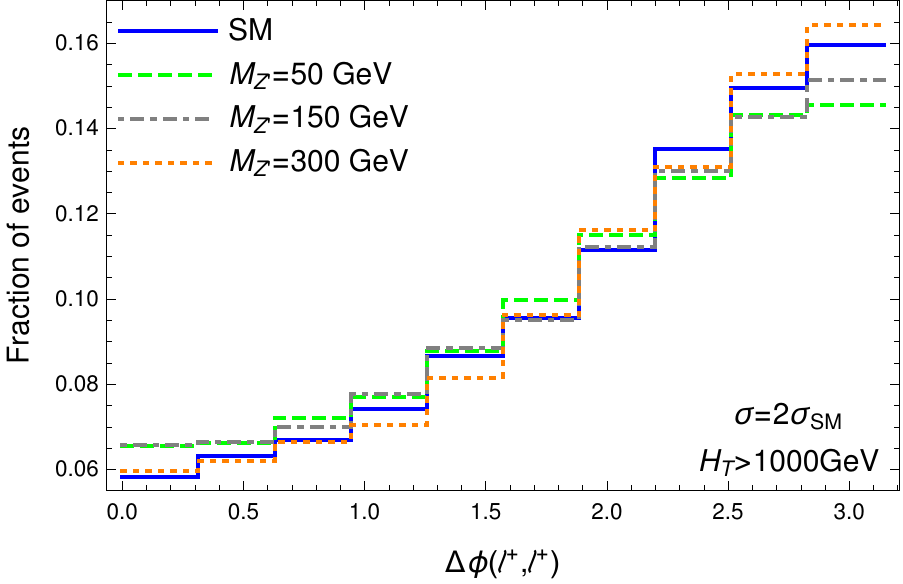}} \\
\caption{\small Distribution of the azimuthal separation between the same-sign leptons,  $\Delta\phi(\ell^+,\ell^+)$, for the 2LSS channel after preselection (see text for details) with 
(left) $\hthad<1000$~GeV and (right) $\hthad>1000$~GeV, normalized to unit area.  
Shown are the predictions for the SM (blue) and the $Z'$ model for (a, b) the tight benchmark points (i.e. giving $\sigma=1.5\sigma_{SM}$) and
(c, d) the loose benchmark points (i.e. giving $\sigma=2\sigma_{SM}$).}
\label{dphizprime}
\end{center}
\end{figure}

\begin{figure}[h!]
\begin{center}
\subfloat[]{\includegraphics[width=0.40\textwidth]{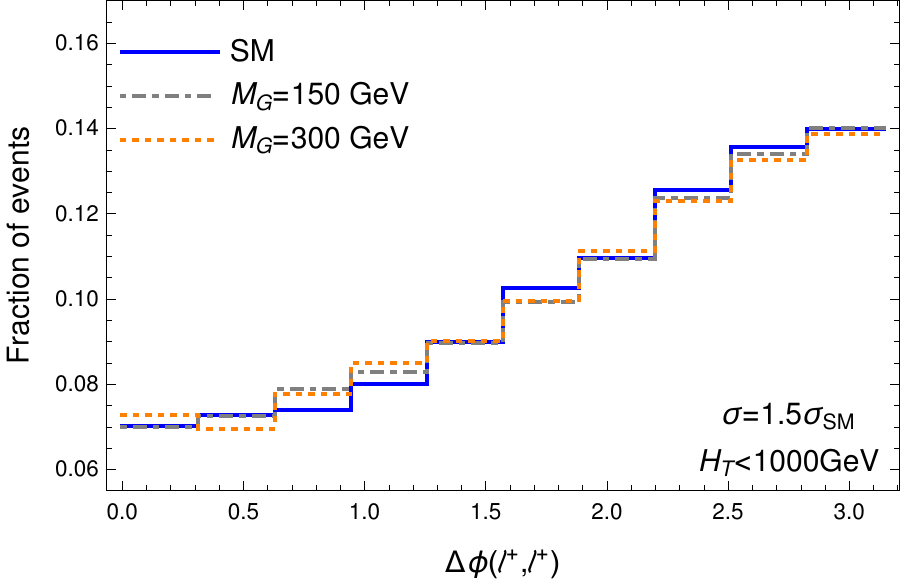}}\hspace{3mm}
\subfloat[]{\includegraphics[width=0.40\textwidth]{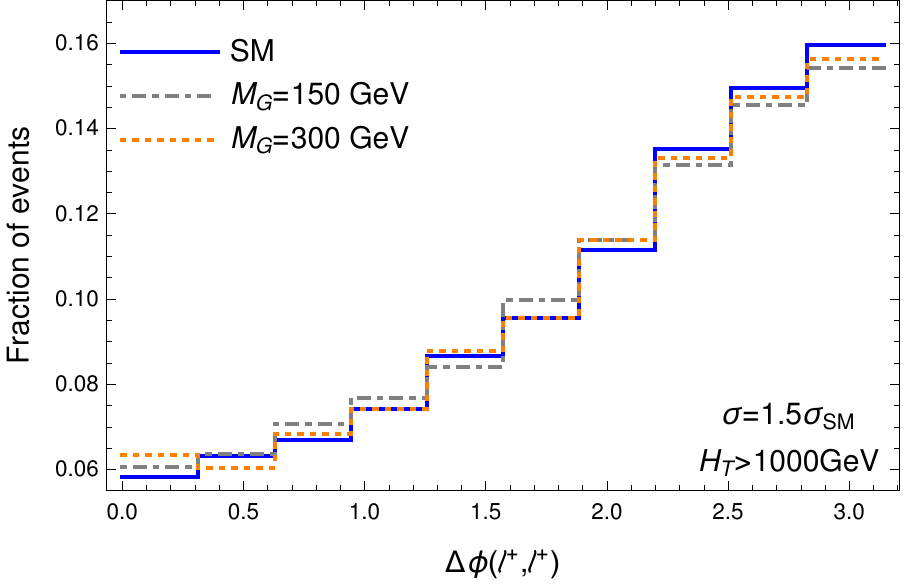}} \\
\subfloat[]{\includegraphics[width=0.40\textwidth]{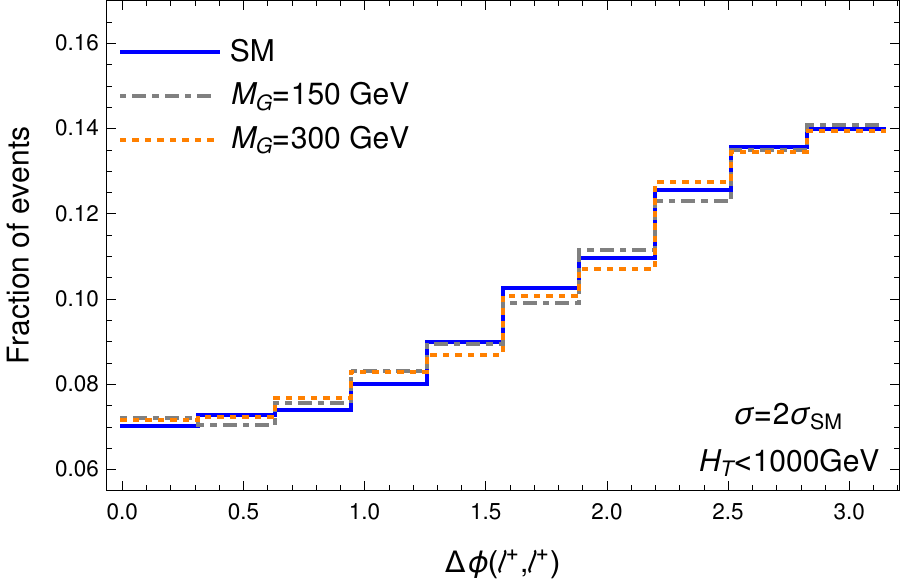}}\hspace{3mm}
\subfloat[]{\includegraphics[width=0.40\textwidth]{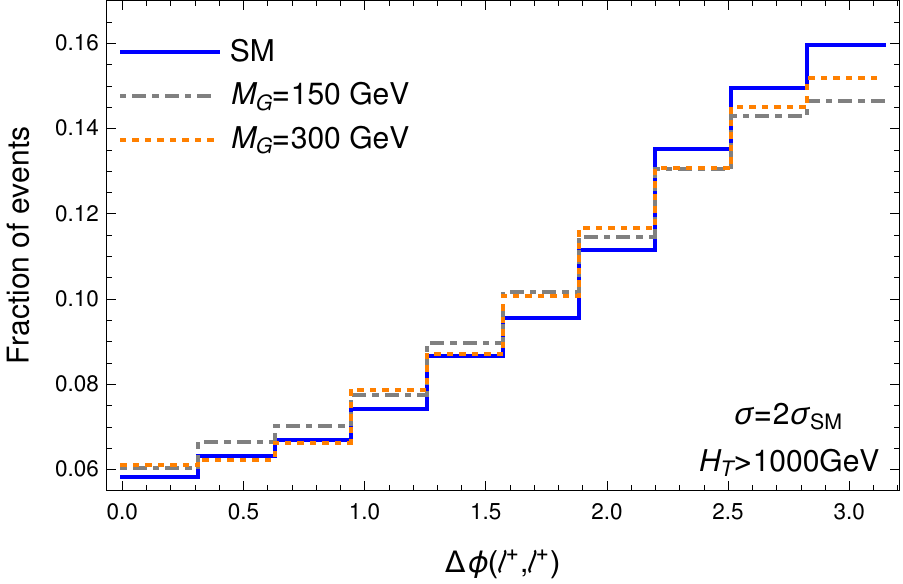}} \\
\caption{\small Distribution of the azimuthal separation between the same-sign leptons,  $\Delta\phi(\ell^+,\ell^+)$, for the 2LSS channel after preselection (see text for details) with 
(left) $\hthad<1000$~GeV and (right) $\hthad>1000$~GeV, normalized to unit area.  
Shown are the predictions for the SM (blue) and the Graviton model for (a, b) the tight benchmark points (i.e. giving $\sigma=1.5\sigma_{SM}$) and
(c, d) the loose benchmark points (i.e. giving $\sigma=2\sigma_{SM}$).}
\label{dphigraviton}
\end{center}
\end{figure}

\setcounter{equation}{0}
We present in this appendix the plots which complement the results in the main body.   Figures \ref{htscalar2}, \ref{Att2}, \ref{dphi-inclusive2}, \ref{dphiscalar}, \ref{dphipseudoscalar}, \ref{dphizprime} and \ref{dphigraviton} contain the same or similar analysis as those presented in text, for all the benchmark points.  In particular, results for loose benchmark points are only presented in this appendix.

\FloatBarrier

\section{Loop functions}
\setcounter{equation}{0}
\label{loop}

We provide some more details about the NP models presented in Section \ref{section:2}.

\noindent{\bf Scalar NP: $\phi$}

\begin{eqnarray}
    {\cal L}_{\phi g g}^\textrm{eff} &=& \frac{\alpha_s}{12\pi}  \frac{g_{\phi t}}{m_t} F(z_t)  G^a_{\mu\nu} G_a^{\mu\nu}  \phi   \\
    {\cal L}_{\phi \gamma \gamma}^\textrm{eff} &=& \frac{2\alpha}{9\pi}  \frac{g_{\phi t}}{m_t}  F_{\mu\nu} F^{\mu\nu}  \phi   \\
F ( z_t )&=& \frac{3}{2} z_t \left(1+(1-z_t) \arcsin ^2 \left[\sqrt{1/z_t} \right] \right)   \\
  z_t &=& (2 m_{t}/M_{\phi})^2  \\
  \Gamma (\phi \rightarrow g g) &=&  \frac{\alpha_s ^2 M_{\phi}^3}{72 \pi ^3}
    \left| \frac{g_{\phi t}}{m_t} F ( z_t )  \right| ^2   \\
\Gamma (\phi \rightarrow \gamma \gamma) &=&  \frac{\alpha ^2 M_{\phi}^3}{81 \pi ^3}
    \left| \frac{g_{\phi t}}{m_t} F ( z_t )  \right| ^2  
\end{eqnarray}

\noindent{\bf Pseudo-scalar NP: $A$}

\begin{eqnarray}
    {\cal L}_{A g g}^\textrm{eff} &=& \frac{\alpha_s}{4\pi}  \frac{g_{A t}}{m_t} H(z_t)  G^a_{\mu\nu} \tilde G_a^{\mu\nu}  A  \\
    {\cal L}_{A \gamma \gamma}^\textrm{eff} &=& \frac{\alpha}{3\pi}  \frac{g_{A t}}{m_t}  F_{\mu\nu} \tilde F^{\mu\nu}  A  \\
H ( z_t )&=&  z_t  \arcsin ^2 \left[\sqrt{1/z_t} \right]  \\
  z_t &=& (2 m_{t}/M_{A})^2  \\
  \Gamma (A \rightarrow g g) &=&  \frac{\alpha_s ^2 M_{\phi}^3}{72 \pi ^3}
    \left| \frac{g_{A t}}{m_t} H ( z_t )  \right| ^2  \\
\Gamma (A \rightarrow \gamma \gamma) &=&  \frac{\alpha ^2 M_{A}^3}{81 \pi ^3}
    \left| \frac{g_{A t}}{m_t} H ( z_t )  \right| ^2 
\end{eqnarray}

\newpage
\noindent{\bf Graviton NP: $G$}

\begin{eqnarray}
    {\cal L}_{Ggg}^{eff}&=& -\frac{\alpha_s}{12\pi \Lambda} g_{G t} A_G(z_t,\mu_0) \hat G_{\mu\nu}  \left(\frac{\eta^{\mu\nu}}{4} \, G_a^{\rho\sigma} G^a_{\rho\sigma}  - G_a^{\mu\rho} G^{a\nu}_\rho  \right) \\
    {\cal L}_{G \gamma \gamma}^\textrm{eff}&=& -\frac{2\alpha}{9\pi \Lambda} g_{G t} A_G(z_t,\mu_0) \hat G_{\mu\nu}  \left( \frac{\eta^{\mu\nu}}{4} \, F^{\rho\sigma} F_{\rho\sigma}  - F^{\mu\rho} F^{\nu}_\rho  \right)  \\
A_G(z_t,\mu_0)&=& -\frac{1}{12}\left[
\frac{9}{4}z_t(z_t+2) [2 \tan^{-1}(\sqrt{z_t-1}) - \pi]^2
\right.
\nonumber \\ &&
\left.
- 3(5z_t+4)\sqrt{z_t-1}[2 \tan^{-1}(\sqrt{z_t-1}) - \pi]
- 39z_t  - 35 -12 \ln\frac{\mu_0^2}{m_t^2} \right]  \\
  z_t &=& (2 m_{t}/M_{G})^2  \\
   \Gamma(G \to gg) &=& \frac{ M_{G}^3}{\pi \Lambda^2} \frac{\alpha_s^2}{1440\pi^2} \left| g_{G t} A_G( z_t, \mu_0 )   \right| ^2\\
   \Gamma(G \rightarrow \gamma \gamma) &=&  \frac{ M_{G}^3}{\pi \Lambda^2} \frac{\alpha^2}{1620\pi^2} \left| g_{G t} A_G( z_t, \mu_0 )   \right| ^2  
\end{eqnarray}

Here $\mu_0$ is the renormalization scale which we have set it to $\mu_0=M_{G}$ \cite{Alvarez:2016ljl} throughout this work.
\section{Simulation details}
\label{appendix}
\setcounter{equation}{0}

Along the article we have used {\tt MadGraph5\_aMC@NLO} \cite{Alwall:2014hca} for matrix level generation, {\tt Pythia} \cite{Sjostrand:2014zea,Sjostrand:2006za} for showering, hadronization and ISR and FSR, and {\tt Delphes} \cite{deFavereau:2013fsa} for detector simulation.  We have used {\tt Madspin} \cite{Artoisenet:2012st,Frixione:2007zp} to decay top quarks while preserving the spin orientation.  The NP models have been implemented through {\tt FeynRules} \cite{Alloul:2013bka}.

In all cases we have included in the simulation QCD and Electroweak leading order effects.  Although Electroweak corrections are a minor correction of the order $\sim 5 \%$ in SM $pp\to t \bar t t \bar t$ production, it can account up to $\sim 30\%$ for some studied NP Benchmark Points.  In general the interference is enhanced with the Electroweak particles of the same Nature as the NP.  For the sake of obtaining reasonable results in a reasonable time we have restricted the proton partons to the gluon, the valence quarks $u$ and $d$, and their anti-particles $\bar u$ and $\bar d$.  This approximation is converted into a difference in cross-section of about 1\%.  The spin-0 and spin-1 NP models contain 524 Feynman diagrams to produce a four-top final state in the aforementioned conditions, whereas spin-2 NP model requires 588 Feynman diagrams.  These extra 64 diagrams are because of the 4-particle vertices needed to conserve gauge invariance in the spin-2 NP Lagrangian.

In Section \ref{constraints} we have simulated SM and NP processes using {\tt MadGraph5\_aMC@NLO} in its original tune.  Since it is the ratio of NP to SM what we compute, we have not required a $k$-factor.  We have not applied cuts on the $pp\to t \bar t t \bar t$ process, whereas for the di-photon generation we have used the same cuts as the cited searches. 

In Section \ref{section:3} we have simulated tops decaying to final state using {\tt Madspin}, {\tt Pythia} and {\tt Delphes}.  All simulations are at leading order using NNPDF30\_lo\_as\_0118 PDF and a k-factor of $1.26$ is extracted from Ref.\cite{Frederix:2017wme,Sirunyan:2019wxt}.  For the sake of computational resources we have only decayed the tops using {\tt Madspin}.  We have set Delphes parameters as in Ref.~\cite{Alvarez:2018gxs} which is tuned for CMS results in Ref.~\cite{Sirunyan:2017roi}.  

Objects at the detector level are defined as follows.  Electrons are required to have $p_{T}>20$ GeV and $\left | \eta \right |<2.5$. Muons are required to have $p_{T}>20$ GeV and $\left | \eta \right |<2.4$.  Hadronically decaying taus $p_{T}>20$ GeV and $\left | \eta \right |<2.5$.  Jets are required to have $p_{T}>40$ GeV and $\left | \eta \right |<2.4$.  Whereas b-tagged jets are demanded to have $p_{T}>25$ GeV and $\left | \eta \right |<2.4$.

\subsection{Computational resources overview}

Four-top is a very populated final state, and with $\sim 500$ Feynman diagrams when the proton PDF is restricted to $g,\, u, \, \bar u, \, d $ and $\bar d$.  If in addition this partonic state is decayed with correlated spins in at least two of the partons, the simulation becomes still more involved.  When simulating only SM the simulation includes QED tree corrections.  When including NP, the Feynman diagrams are increased because of new diagrams. The following table is a representative sample of the computational resources used to simulate some of the results in the manuscript.

{\small
\begin{center}
\begin{tabular}{|p{.195\linewidth}|p{.15\linewidth}|p{.19\linewidth}|p{.27\linewidth}|p{.065\textwidth}|}
      	\hline
      	\textbf{Madgraph} & \textbf{Model} & \textbf{CPU \newline specifications} & \textbf{Comments} & \textbf{CPU \newline hours}\\
	\hline
        $pp \to t \bar t t \bar t$ QED=2 \newline adapted for like-helicity tops & Scalar/\newline Pseudo-scalar & Intel Quad 2.4GHz & Fig.~\ref{Att}a--d. Grid of 4 BPs with 1M events each. \newline Total: 4M events & 100 \\
	\hline
	$pp \to t \bar t t \bar t$ QED=2 $\newline+$Madspin$+$Pythia $+$Delphes  & Scalar/\newline Pseudo-scalar/\newline Vector Z' & Intel i7 3.4GHz & Complete calculation for each BP in Figs.~\ref{htscalar}, \ref{dphi-inclusive}, \ref{dphiscalar}--\ref{dphizprime}. \newline Total: 0.5M events & 200 \\
      	\hline	
       	 $pp \to t \bar t t \bar t$ QED=2  & Graviton & Intel i7 4GHz & Fig.~\ref{bounds1}d four-top contour levels. Grid of 170 BPs with 10k events each.  \newline Total: 1.7M events & 1400 \\
      	\hline	
     	\end{tabular}
  \end{center}

\newpage

\bibliographystyle{JHEP}
\bibliography{biblio}

\providecommand{\href}[2]{#2}\begingroup\raggedright\begin{thebibliography}{10}

\bibitem{Aad:2012tfa}
{ATLAS Collaboration}, \emph{{Observation of a new particle in the search for
  the Standard Model Higgs boson with the ATLAS detector at the LHC}},
  \href{http://dx.doi.org/10.1016/j.physletb.2012.08.020}{\emph{Phys. Lett. B}
  {\bf 716} (2012) 1}, [\href{http://arxiv.org/abs/arXiv:1207.7214}{{\tt
  arXiv:1207.7214}}].

\bibitem{Chatrchyan:2012xdj}
{CMS Collaboration}, \emph{{Observation of a New Boson at a Mass of 125 GeV
  with the CMS Experiment at the LHC}},
  \href{http://dx.doi.org/10.1016/j.physletb.2012.08.021}{\emph{Phys. Lett. B}
  {\bf 716} (2012) 30}, [\href{http://arxiv.org/abs/arXiv:1207.7235}{{\tt
  arXiv:1207.7235}}].

\bibitem{DiMicco:2019ngk}
J.~Alison et~al., \emph{{Higgs boson pair production at colliders: status and
  perspectives}},  in \emph{{Double Higgs Production at Colliders Batavia, IL,
  USA, September 4, 2018-9, 2019}} (B.~Di~Micco, M.~Gouzevitch, J.~Mazzitelli
  and C.~Vernieri, eds.), 2019.
\newblock \href{http://arxiv.org/abs/arXiv:1910.00012}{{\tt arXiv:1910.00012}}.

\bibitem{Aad:2019yxi}
{ATLAS Collaboration}, \emph{{Search for non-resonant Higgs boson pair
  production in the $bb\ell\nu\ell\nu$ final state with the ATLAS detector in
  $pp$ collisions at $\sqrt{s} = 13$ TeV}},
  \href{http://arxiv.org/abs/arXiv:1908.06765}{{\tt arXiv:1908.06765}}.

\bibitem{CMS:2019vgr}
{CMS Collaboration}, \emph{{Search for the resonant production of a pair of
  Higgs bosons decaying to the $b\bar{b}ZZ$ final state}},
  {\emph{CMS-PAS-HIG-18-013} (2019) }.
  \url{https://cds.cern.ch/record/2682621}.

\bibitem{CMS:2018rbc}
{CMS Collaboration}, \emph{{Measurement of the associated production of a Higgs
  boson and a pair of top-antitop quarks with the Higgs boson decaying to two
  photons in proton-proton collisions at $\sqrt{s}=13~\mathrm{TeV}$}},
  {\emph{CMS-PAS-HIG-18-018} (2018) }.
  \url{https://cds.cern.ch/record/2649208}.

\bibitem{Madaffari:2018bbq}
D.~Madaffari, \emph{{Higgs boson production in association with a top quark
  pair at the LHC}}, \href{http://dx.doi.org/10.22323/1.330.0019}{\emph{PoS}
  {\bf ALPS2018} (2018) 019}.

\bibitem{Aaboud:2018urx}
{ATLAS Collaboration}, \emph{{Observation of Higgs boson production in
  association with a top quark pair at the LHC with the ATLAS detector}},
  \href{http://dx.doi.org/10.1016/j.physletb.2018.07.035}{\emph{Phys. Lett. B}
  {\bf 784} (2018) 173}, [\href{http://arxiv.org/abs/arXiv:1806.00425}{{\tt
  arXiv:1806.00425}}].

\bibitem{Aad:2015kqa}
{ATLAS Collaboration}, \emph{{Search for production of vector-like quark pairs
  and of four top quarks in the lepton-plus-jets final state in $pp$ collisions
  at $\sqrt{s}=8$~TeV with the ATLAS detector}},
  \href{http://dx.doi.org/10.1007/JHEP08(2015)105}{\emph{JHEP} {\bf 08} (2015)
  105}, [\href{http://arxiv.org/abs/arXiv:1505.04306}{{\tt arXiv:1505.04306}}].

\bibitem{Aad:2015gdg}
{ATLAS Collaboration}, \emph{{Analysis of events with $b$-jets and a pair of
  leptons of the same charge in $pp$ collisions at $\sqrt{s}=8$~TeV with the
  ATLAS detector}},
  \href{http://dx.doi.org/10.1007/JHEP10(2015)150}{\emph{JHEP} {\bf 10} (2015)
  150}, [\href{http://arxiv.org/abs/arXiv:1504.04605}{{\tt arXiv:1504.04605}}].

\bibitem{Sirunyan:2019wxt}
{CMS Collaboration}, \emph{{Search for production of four top quarks in final
  states with same-sign or multiple leptons in proton-proton collisions at
  $\sqrt{s}=13$~TeV}},  \href{http://arxiv.org/abs/arXiv:1908.06463}{{\tt
  arXiv:1908.06463}}.

\bibitem{Cao:2016wib}
Q.-H. Cao, S.-L. Chen and Y.~Liu, \emph{{Probing Higgs Width and Top Quark
  Yukawa Coupling from $t\bar{t}H$ and $t\bar{t}t\bar{t}$ Productions}},
  \href{http://dx.doi.org/10.1103/PhysRevD.95.053004}{\emph{Phys. Rev. D} {\bf
  95} (2017) 053004}, [\href{http://arxiv.org/abs/arXiv:1602.01934}{{\tt
  arXiv:1602.01934}}].

\bibitem{Cao:2019ygh}
Q.-H. Cao, S.-L. Chen, Y.~Liu, R.~Zhang and Y.~Zhang, \emph{{Limiting top
  quark-Higgs boson interaction and Higgs-boson width from multitop
  productions}},
  \href{http://dx.doi.org/10.1103/PhysRevD.99.113003}{\emph{Phys. Rev. D} {\bf
  99} (2019) 113003}, [\href{http://arxiv.org/abs/arXiv:1901.04567}{{\tt
  arXiv:1901.04567}}].

\bibitem{Englert:2019zmt}
C.~Englert, G.~F. Giudice, A.~Greljo and M.~Mccullough, \emph{{The
  $\hat{H}$-Parameter: An Oblique Higgs View}},
  \href{http://dx.doi.org/10.1007/JHEP09(2019)041}{\emph{JHEP} {\bf 09} (2019)
  041}, [\href{http://arxiv.org/abs/arXiv:1903.07725}{{\tt arXiv:1903.07725}}].

\bibitem{Alvarez:2016nrz}
E.~Alvarez, D.~A. Faroughy, J.~F. Kamenik, R.~Morales and A.~Szynkman,
  \emph{{Four Tops for LHC}},
  \href{http://dx.doi.org/10.1016/j.nuclphysb.2016.11.024}{\emph{Nucl. Phys. B}
  {\bf 915} (2017) 19}, [\href{http://arxiv.org/abs/arXiv:1611.05032}{{\tt
  arXiv:1611.05032}}].

\bibitem{Battaglia:2010xq}
M.~Battaglia and G.~Servant, \emph{{Four-top production and t tbar + missing
  energy events at multi TeV e+e- colliders}},
  \href{http://dx.doi.org/10.1393/ncc/i2010-10611-4}{\emph{Nuovo Cim.} {\bf
  C033N2} (2010) 203}, [\href{http://arxiv.org/abs/arXiv:1005.4632}{{\tt
  arXiv:1005.4632}}].

\bibitem{Degrande:2010kt}
C.~Degrande, J.-M. Gerard, C.~Grojean, F.~Maltoni and G.~Servant,
  \emph{{Non-resonant New Physics in Top Pair Production at Hadron Colliders}},
  \href{http://dx.doi.org/10.1007/JHEP03(2011)125}{\emph{JHEP} {\bf 03} (2011)
  125}, [\href{http://arxiv.org/abs/arXiv:1010.6304}{{\tt arXiv:1010.6304}}].

\bibitem{Alvarez:2017wwr}
E.~Alvarez and M.~Estevez, \emph{{$t \bar t b \bar b$ as a probe of new physics
  at the LHC}}, \href{http://dx.doi.org/10.1103/PhysRevD.96.035016}{\emph{Phys.
  Rev. D} {\bf 96} (2017) 035016},
  [\href{http://arxiv.org/abs/arXiv:1701.04427}{{\tt arXiv:1701.04427}}].

\bibitem{Lillie:2007hd}
B.~Lillie, J.~Shu and T.~M.~P. Tait, \emph{{Top Compositeness at the Tevatron
  and LHC}}, \href{http://dx.doi.org/10.1088/1126-6708/2008/04/087}{\emph{JHEP}
  {\bf 04} (2008) 087}, [\href{http://arxiv.org/abs/arXiv:0712.3057}{{\tt
  arXiv:0712.3057}}].

\bibitem{Greiner:2014qna}
N.~Greiner, K.~Kong, J.-C. Park, S.~C. Park and J.-C. Winter,
  \emph{{Model-Independent Production of a Top-Philic Resonance at the LHC}},
  \href{http://dx.doi.org/10.1007/JHEP04(2015)029}{\emph{JHEP} {\bf 04} (2015)
  029}, [\href{http://arxiv.org/abs/arXiv:1410.6099}{{\tt arXiv:1410.6099}}].

\bibitem{Azzi:2019yne}
{\scshape HL-LHC, HE-LHC Working Group} collaboration, P.~Azzi et~al.,
  \emph{{Standard Model Physics at the HL-LHC and HE-LHC}},
  \href{http://arxiv.org/abs/arXiv:1902.04070}{{\tt arXiv:1902.04070}}.

\bibitem{Calvet:2012rk}
S.~Calvet, B.~Fuks, P.~Gris and L.~Valery, \emph{{Searching for sgluons in
  multitop events at a center-of-mass energy of 8 TeV}},
  \href{http://dx.doi.org/10.1007/JHEP04(2013)043}{\emph{JHEP} {\bf 04} (2013)
  043}, [\href{http://arxiv.org/abs/arXiv:1212.3360}{{\tt arXiv:1212.3360}}].

\bibitem{Darme:2018dvz}
L.~Darmé, B.~Fuks and M.~Goodsell, \emph{{Cornering sgluons with
  four-top-quark events}},
  \href{http://dx.doi.org/10.1016/j.physletb.2018.08.001}{\emph{Phys. Lett. B}
  {\bf 784} (2018) 223}, [\href{http://arxiv.org/abs/arXiv:1805.10835}{{\tt
  arXiv:1805.10835}}].

\bibitem{Zhang:2017mls}
C.~Zhang, \emph{{Constraining $qqtt$ operators from four-top production: a case
  for enhanced EFT sensitivity}},
  \href{http://dx.doi.org/10.1088/1674-1137/42/2/023104}{\emph{Chin. Phys. C}
  {\bf 42} (2018) 023104}, [\href{http://arxiv.org/abs/arXiv:1708.05928}{{\tt
  arXiv:1708.05928}}].

\bibitem{Alvarez:2019knh}
E.~Alvarez, F.~Lamagna and M.~Szewc, \emph{{Topic Model for four-top at the
  LHC}},  \href{http://arxiv.org/abs/1911.09699}{{\tt 1911.09699}}.

\bibitem{Aad:2014pda}
{ATLAS Collaboration}, \emph{{Search for supersymmetry at $\sqrt{s}=8$~TeV in
  final states with jets and two same-sign leptons or three leptons with the
  ATLAS detector}},
  \href{http://dx.doi.org/10.1007/JHEP06(2014)035}{\emph{JHEP} {\bf 06} (2014)
  035}, [\href{http://arxiv.org/abs/arXiv:1404.2500}{{\tt arXiv:1404.2500}}].

\bibitem{Khachatryan:2014sca}
{CMS Collaboration}, \emph{{Search for Standard Model Production of Four Top
  Quarks in the Lepton + Jets Channel in pp Collisions at $\sqrt{s}=8$~TeV}},
  \href{http://dx.doi.org/10.1007/JHEP11(2014)154}{\emph{JHEP} {\bf 11} (2014)
  154}, [\href{http://arxiv.org/abs/arXiv:1409.7339}{{\tt arXiv:1409.7339}}].

\bibitem{Sirunyan:2017tep}
{CMS Collaboration}, \emph{{Search for standard model production of four top
  quarks in proton-proton collisions at $\sqrt{s}=13$~TeV}},
  \href{http://dx.doi.org/10.1016/j.physletb.2017.06.064}{\emph{Phys. Lett. B}
  {\bf 772} (2017) 336}, [\href{http://arxiv.org/abs/arXiv:1702.06164}{{\tt
  arXiv:1702.06164}}].

\bibitem{Sirunyan:2017uyt}
{CMS Collaboration}, \emph{{Search for physics beyond the standard model in
  events with two leptons of same sign, missing transverse momentum, and jets
  in proton-proton collisions at $\sqrt{s}=13$~TeV}},
  \href{http://dx.doi.org/10.1140/epjc/s10052-017-5079-z}{\emph{Eur. Phys. J.
  C} {\bf 77} (2017) 578}, [\href{http://arxiv.org/abs/arXiv:1704.07323}{{\tt
  arXiv:1704.07323}}].

\bibitem{Sirunyan:2017roi}
{CMS Collaboration}, \emph{{Search for standard model production of four top
  quarks with same-sign and multilepton final states in proton-proton
  collisions at $\sqrt{s}=13$~TeV}},
  \href{http://dx.doi.org/10.1140/epjc/s10052-018-5607-5}{\emph{Eur. Phys. J.
  C} {\bf 78} (2018) 140}, [\href{http://arxiv.org/abs/arXiv:1710.10614}{{\tt
  arXiv:1710.10614}}].

\bibitem{Aaboud:2018jsj}
{ATLAS Collaboration}, \emph{{Search for four-top-quark production in the
  single-lepton and opposite-sign dilepton final states in pp collisions at
  $\sqrt{s}=13$~TeV with the ATLAS detector}},
  \href{http://dx.doi.org/10.1103/PhysRevD.99.052009}{\emph{Phys. Rev. D} {\bf
  99} (2019) 052009}, [\href{http://arxiv.org/abs/arXiv:1811.02305}{{\tt
  arXiv:1811.02305}}].

\bibitem{Aaboud:2018xpj}
{ATLAS Collaboration}, \emph{{Search for new phenomena in events with
  same-charge leptons and $b$-jets in $pp$ collisions at $\sqrt{s}=13$~TeV with
  the ATLAS detector}},
  \href{http://dx.doi.org/10.1007/JHEP12(2018)039}{\emph{JHEP} {\bf 12} (2018)
  039}, [\href{http://arxiv.org/abs/arXiv:1807.11883}{{\tt arXiv:1807.11883}}].

\bibitem{Sirunyan:2019nxl}
{CMS Collaboration}, \emph{{Search for the Production of Four Top Quarks in the
  Single-Lepton and Opposite-Sign Dilepton Final States in Proton-Proton
  Collisions at $\sqrt{s}=13$~TeV}},
  \href{http://arxiv.org/abs/arXiv:1906.02805}{{\tt arXiv:1906.02805}}.

\bibitem{Frederix:2017wme}
R.~Frederix, D.~Pagani and M.~Zaro, \emph{{Large NLO corrections in
  $t\bar{t}W^{\pm}$ and $t\bar{t}t\bar{t}$ hadroproduction from supposedly
  subleading EW contributions}},
  \href{http://dx.doi.org/10.1007/JHEP02(2018)031}{\emph{JHEP} {\bf 02} (2018)
  031}, [\href{http://arxiv.org/abs/arXiv:1711.02116}{{\tt arXiv:1711.02116}}].

\bibitem{Contino:2006nn}
R.~Contino, T.~Kramer, M.~Son and R.~Sundrum, \emph{{Warped/composite
  phenomenology simplified}},
  \href{http://dx.doi.org/10.1088/1126-6708/2007/05/074}{\emph{JHEP} {\bf 05}
  (2007) 074}, [\href{http://arxiv.org/abs/arXiv:hep-ph/0612180}{{\tt
  arXiv:hep-ph/0612180}}].

\bibitem{Marzocca:2012zn}
D.~Marzocca, M.~Serone and J.~Shu, \emph{{General Composite Higgs Models}},
  \href{http://dx.doi.org/10.1007/JHEP08(2012)013}{\emph{JHEP} {\bf 08} (2012)
  013}, [\href{http://arxiv.org/abs/arXiv:1205.0770}{{\tt arXiv:1205.0770}}].

\bibitem{Caracciolo:2012je}
F.~Caracciolo, A.~Parolini and M.~Serone, \emph{{UV Completions of Composite
  Higgs Models with Partial Compositeness}},
  \href{http://dx.doi.org/10.1007/JHEP02(2013)066}{\emph{JHEP} {\bf 02} (2013)
  066}, [\href{http://arxiv.org/abs/arXiv:1211.7290}{{\tt arXiv:1211.7290}}].

\bibitem{Dev:2014yca}
P.~S. Bhupal~Dev and A.~Pilaftsis, \emph{{Maximally Symmetric Two Higgs Doublet
  Model with Natural Standard Model Alignment}},
  \href{http://dx.doi.org/10.1007/JHEP11(2015)147,
  10.1007/JHEP12(2014)024}{\emph{JHEP} {\bf 12} (2014) 024},
  [\href{http://arxiv.org/abs/arXiv:1408.3405}{{\tt arXiv:1408.3405}}].
  [Erratum: JHEP \textbf{11} (2015) 147].

\bibitem{Landau:1948kw}
L.~D. Landau, \emph{{On the angular momentum of a system of two photons}},
  \href{http://dx.doi.org/10.1016/B978-0-08-010586-4.50070-5}{\emph{Dokl. Akad.
  Nauk Ser. Fiz.} {\bf 60} (1948) 207}.

\bibitem{Yang:1950rg}
C.-N. Yang, \emph{{Selection Rules for the Dematerialization of a Particle Into
  Two Photons}}, \href{http://dx.doi.org/10.1103/PhysRev.77.242}{\emph{Phys.
  Rev.} {\bf 77} (1950) 242}.

\bibitem{Beenakker:2015mra}
W.~Beenakker, R.~Kleiss and G.~Lustermans, \emph{{No Landau-Yang in QCD}},
  \href{http://arxiv.org/abs/arXiv:1508.07115}{{\tt arXiv:1508.07115}}.

\bibitem{Pleitez:2018lct}
V.~Pleitez, \emph{{Angular momentum and parity of a two gluon system}},
  \href{http://arxiv.org/abs/arXiv:1801.09294}{{\tt arXiv:1801.09294}}.

\bibitem{Cacciari:2015ela}
M.~Cacciari, L.~Del~Debbio, J.~R. Espinosa, A.~D. Polosa and M.~Testa, \emph{{A
  note on the fate of the Landau–Yang theorem in non-Abelian gauge
  theories}},
  \href{http://dx.doi.org/10.1016/j.physletb.2015.12.053}{\emph{Phys. Lett. B}
  {\bf 753} (2016) 476}, [\href{http://arxiv.org/abs/arXiv:1509.07853}{{\tt
  arXiv:1509.07853}}].

\bibitem{Pleitez:2015cpa}
V.~Pleitez, \emph{{The angular momentum of two massless fields revisited}},
  \href{http://arxiv.org/abs/arXiv:1508.01394}{{\tt arXiv:1508.01394}}.

\bibitem{Alvarez:2016ljl}
E.~Alvarez, L.~Da~Rold, J.~Mazzitelli and A.~Szynkman, \emph{{Graviton
  resonance phenomenology and a pseudo-Nambu-Goldstone boson Higgs at the
  LHC}}, \href{http://dx.doi.org/10.1103/PhysRevD.95.115012}{\emph{Phys. Rev.
  D} {\bf 95} (2017) 115012},
  [\href{http://arxiv.org/abs/arXiv:1610.08451}{{\tt arXiv:1610.08451}}].

\bibitem{Kuhn:2013zoa}
J.~H. Kühn, A.~Scharf and P.~Uwer, \emph{{Weak Interactions in Top-Quark Pair
  Production at Hadron Colliders: An Update}},
  \href{http://dx.doi.org/10.1103/PhysRevD.91.014020}{\emph{Phys. Rev. D} {\bf
  91} (2015) 014020}, [\href{http://arxiv.org/abs/arXiv:1305.5773}{{\tt
  arXiv:1305.5773}}].

\bibitem{CMS:2019unu}
{CMS Collaboration}, \emph{{Constraining the top quark Yukawa coupling from
  $t\bar{t}$ differential cross sections in the lepton+jets final state in
  proton-proton collisions at $\sqrt{s}=13$~TeV}}, {\emph{CMS-PAS-TOP-17-004}
  (2019) }. \url{https://cds.cern.ch/record/2665937}.

\bibitem{Aad:2014ioa}
{ATLAS Collaboration}, \emph{{Search for Scalar Diphoton Resonances in the Mass
  Range $65-600$ GeV with the ATLAS Detector in $pp$ Collision Data at
  $\sqrt{s}$ = 8 $TeV$}},
  \href{http://dx.doi.org/10.1103/PhysRevLett.113.171801}{\emph{Phys. Rev.
  Lett.} {\bf 113} (2014) 171801},
  [\href{http://arxiv.org/abs/arXiv:1407.6583}{{\tt arXiv:1407.6583}}].

\bibitem{Aaboud:2017yyg}
{ATLAS Collaboration}, \emph{{Search for new phenomena in high-mass diphoton
  final states using 37 fb$^{-1}$ of proton--proton collisions collected at
  $\sqrt{s}=13$ TeV with the ATLAS detector}},
  \href{http://dx.doi.org/10.1016/j.physletb.2017.10.039}{\emph{Phys. Lett. B}
  {\bf 775} (2017) 105}, [\href{http://arxiv.org/abs/arXiv:1707.04147}{{\tt
  arXiv:1707.04147}}].

\bibitem{ATLAS:2018xad}
{ATLAS Collaboration}, \emph{{Search for resonances in the 65 to 110 GeV
  diphoton invariant mass range using 80 fb$^{-1}$ of $pp$ collisions collected
  at $\sqrt{s}=13$ TeV with the ATLAS detector}}, {\emph{ATLAS-CONF-2018-025}
  (2018) }. \url{https://cds.cern.ch/record/2628760}.

\bibitem{Khachatryan:2015qba}
{CMS Collaboration}, \emph{{Search for diphoton resonances in the mass range
  from 150 to 850 GeV in pp collisions at $\sqrt{s} =$ 8 TeV}},
  \href{http://dx.doi.org/10.1016/j.physletb.2015.09.062}{\emph{Phys. Lett. B}
  {\bf 750} (2015) 494}, [\href{http://arxiv.org/abs/arXiv:1506.02301}{{\tt
  arXiv:1506.02301}}].

\bibitem{Sirunyan:2018aui}
{CMS Collaboration}, \emph{{Search for a standard model-like Higgs boson in the
  mass range between 70 and 110 GeV in the diphoton final state in
  proton-proton collisions at $\sqrt{s}=$ 8 and 13 TeV}},
  \href{http://dx.doi.org/10.1016/j.physletb.2019.03.064}{\emph{Phys. Lett. B}
  {\bf 793} (2019) 320}, [\href{http://arxiv.org/abs/arXiv:1811.08459}{{\tt
  arXiv:1811.08459}}].

\bibitem{Bevilacqua:2012em}
G.~Bevilacqua and M.~Worek, \emph{{Constraining BSM Physics at the LHC: Four
  top final states with NLO accuracy in perturbative QCD}},
  \href{http://dx.doi.org/10.1007/JHEP07(2012)111}{\emph{JHEP} {\bf 07} (2012)
  111}, [\href{http://arxiv.org/abs/arXiv:1206.3064}{{\tt arXiv:1206.3064}}].

\bibitem{vanderBij:1988ac}
J.~J. van~der Bij and E.~W.~N. Glover, \emph{{$Z$ Boson Production and Decay
  via Gluons}},
  \href{http://dx.doi.org/10.1016/0550-3213(89)90317-9}{\emph{Nucl. Phys. B}
  {\bf 313} (1989) 237}.

\bibitem{Alwall:2014hca}
J.~Alwall, R.~Frederix, S.~Frixione, V.~Hirschi, F.~Maltoni, O.~Mattelaer
  et~al., \emph{{The automated computation of tree-level and next-to-leading
  order differential cross sections, and their matching to parton shower
  simulations}}, \href{http://dx.doi.org/10.1007/JHEP07(2014)079}{\emph{JHEP}
  {\bf 07} (2014) 079}, [\href{http://arxiv.org/abs/arXiv:1405.0301}{{\tt
  arXiv:1405.0301}}].

\bibitem{Sjostrand:2014zea}
T.~Sj{\"o}strand, S.~Ask, J.~R. Christiansen, R.~Corke, N.~Desai, P.~Ilten
  et~al., \emph{{An Introduction to PYTHIA 8.2}},
  \href{http://dx.doi.org/10.1016/j.cpc.2015.01.024}{\emph{Comput. Phys.
  Commun.} {\bf 191} (2015) 159},
  [\href{http://arxiv.org/abs/arXiv:1410.3012}{{\tt arXiv:1410.3012}}].

\bibitem{Sjostrand:2006za}
T.~Sj{\"o}strand, S.~Mrenna and P.~Z. Skands, \emph{{PYTHIA 6.4 Physics and
  Manual}}, \href{http://dx.doi.org/10.1088/1126-6708/2006/05/026}{\emph{JHEP}
  {\bf 05} (2006) 026}, [\href{http://arxiv.org/abs/arXiv:hep-ph/0603175}{{\tt
  arXiv:hep-ph/0603175}}].

\bibitem{deFavereau:2013fsa}
J.~de~Favereau, C.~Delaere, P.~Demin, A.~Giammanco, V.~Lema{\^\i}tre,
  A.~Mertens et~al., \emph{{DELPHES 3, A modular framework for fast simulation
  of a generic collider experiment}},
  \href{http://dx.doi.org/10.1007/JHEP02(2014)057}{\emph{JHEP} {\bf 02} (2014)
  057}, [\href{http://arxiv.org/abs/1307.6346}{{\tt 1307.6346}}].

\bibitem{Artoisenet:2012st}
P.~Artoisenet, R.~Frederix, O.~Mattelaer and R.~Rietkerk, \emph{{Automatic
  spin-entangled decays of heavy resonances in Monte Carlo simulations}},
  \href{http://dx.doi.org/10.1007/JHEP03(2013)015}{\emph{JHEP} {\bf 03} (2013)
  015}, [\href{http://arxiv.org/abs/arXiv:1212.3460}{{\tt arXiv:1212.3460}}].

\bibitem{Frixione:2007zp}
S.~Frixione, E.~Laenen, P.~Motylinski and B.~R. Webber, \emph{{Angular
  correlations of lepton pairs from vector boson and top quark decays in Monte
  Carlo simulations}},
  \href{http://dx.doi.org/10.1088/1126-6708/2007/04/081}{\emph{JHEP} {\bf 04}
  (2007) 081}, [\href{http://arxiv.org/abs/arXiv:hep-ph/0702198}{{\tt
  arXiv:hep-ph/0702198}}].

\bibitem{Alloul:2013bka}
A.~Alloul, N.~D. Christensen, C.~Degrande, C.~Duhr and B.~Fuks,
  \emph{{FeynRules 2.0 - A complete toolbox for tree-level phenomenology}},
  \href{http://dx.doi.org/10.1016/j.cpc.2014.04.012}{\emph{Comput. Phys.
  Commun.} {\bf 185} (2014) 2250},
  [\href{http://arxiv.org/abs/arXiv:1310.1921}{{\tt arXiv:1310.1921}}].

\bibitem{Alvarez:2018gxs}
E.~Alvarez, L.~Da~Rold, A.~Juste, M.~Szewc and T.~Vazquez~Schroeder, \emph{{A
  composite pNGB leptoquark at the LHC}},
  \href{http://dx.doi.org/10.1007/JHEP12(2018)027}{\emph{JHEP} {\bf 12} (2018)
  027}, [\href{http://arxiv.org/abs/arXiv:1808.02063}{{\tt arXiv:1808.02063}}].

\end{thebibliography}\endgroup

\end{document}